\documentclass[useAMS, usenatbib]{mn2e}

\newcounter{footnew}
\newcommand{\startfoot}{\setcounter{footnew}{0}}
\newcommand{\valfoot}{\stepcounter{footnew}\footnotemark[\value{footnew}]}
\newcommand{\newfoot}{
\valfoot
\renewcommand*{\thefootnote}{\arabic{footnote}}
}
\newcommand{\textfoot}[1]{\newfoot{} {\footnotesize #1 }\\}

\usepackage{natbib}
\usepackage{graphicx}
\usepackage{amssymb}
\usepackage{slantsc}
\usepackage{tablefootnote}
\usepackage{pdflscape}
\usepackage{longtable}
\usepackage{journals}
\usepackage{amsmath}
\usepackage{color}
\usepackage{supertabular,booktabs}
\usepackage[titletoc]{appendix}

\newcommand{\caps}[1]{{\scshape{#1}}}
\newcommand{\figbox}[1]{\makebox[0.5\textwidth]{\includegraphics[width=0.55\textwidth,clip, trim=0.0cm 0.0cm 0.0cm 1cm]{#1}}} % withoutframe
\newcommand{\twofigbox}[2]{\figbox{#1}\hspace{\stretch{0}}\figbox{#2}\\ \vspace{-6pt}}

\begin{document}

\title[]{Mid-UV Studies of the Transitional Millisecond Pulsars XSS J12270$-$4859 and PSR J1023+0038 During Their Radio Pulsar 
States\thanks{Based on proprietary and archival observations with the NASA/ESA Hubble Space Telescope, 
obtained at the Space Telescope Science Institute, which is operated by AURA, Inc., under NASA contract NAS $5-26555$.}}
\author[L.E. Rivera Sandoval et al.]{L. E. Rivera Sandoval$^{1}$\thanks{E-mail: liliana.rivera@ttu.edu}, J. V. Hern\'andez Santisteban$^{1}$, N. Degenaar$^{1}$, 
\newauthor R. Wijnands$^{1}$, C. Knigge$^{2}$, J. M. Miller$^{3}$, M. Reynolds$^{3}$, D. Altamirano$^{2}$, 
\newauthor M. van den Berg$^{4,1}$ and A. Hill$^{5,2}$\\
\\
$^{1}$ Anton Pannekoek Institute for Astronomy, University of Amsterdam, Science Park 904, 1098 XH Amsterdam, The Netherlands\\
$^{2}$ Physics and Astronomy, University of Southampton, Southampton SO17 1BJ, UK\\
$^{3}$ Department of Astronomy, University of Michigan, 1085 South University Ave, Ann Arbor, MI 48109-1107, USA\\
$^{4}$ Harvard-Smithsonian Center for Astrophysics, 60 Garden Street, Cambridge, MA 02138, USA\\
$^{5}$ HAL24K Data Intelligence Labs, Barbara Strozzilaan, 1083 HN Amsterdam, The Netherlands\\}

\pagerange{\pageref{firstpage}--\pageref{lastpage}} \pubyear{2002}

\maketitle

\label{firstpage}

\begin{abstract}
We report mid-UV (MUV) observations taken with \textit{HST}/WFC3, \textit{Swift}/UVOT and \textit{GALEX}/NUV
of the transitional millisecond pulsars XSS J12270$-$4859 and PSR J1023+0038 during their radio pulsar states.
Both systems were detected in our images and showed MUV variability. 
At similar orbital phases, the MUV luminosities of both pulsars are comparable.
This suggests that the emission processes involved in both objects are similar.
We estimated limits on the mass ratio, companion's temperature, inclination, and distance to XSS J12270$-$4859
by using a Markov Chain Monte Carlo algorithm to fit published folded optical light curves.
Using the resulting parameters, we modeled MUV light curves in our \textit{HST} filters. 
The resulting models failed to fit our MUV observations.
Fixing the mass ratio of XSS J12270$-$4859 to the value reported in other studies, we obtained a distance of $\sim3.2$ kpc. This is larger than the one derived from dispersion measure ($\sim1.4$ kpc). 
Assuming a uniform prior for the mass ratio, 
the distance is similar to that from radio measurements.
However, it requires an undermassive companion ($\sim0.01 M_{\sun}$).
We conclude that a direct heating model alone can not fully explain the observations in optical and MUV. 
Therefore, an additional radiation source is needed. 
The source could be an intrabinary shock which contributes to the MUV flux and likely to the optical one as well. 
During the radio pulsar state, the MUV orbital variations of PSR J1023+0038 detected with \textit{GALEX}, suggests the presence of an asymmetric intrabinary shock.

\end{abstract}

\begin{keywords}
X-ray binaries - millisecond pulsars - pulsars: individual: XSS J12270$-$4859  - pulsars: individual: PSR J1023+0038
\end{keywords}

\stepcounter{footnote}

\section{Introduction}

Transitional millisecond pulsars (tMSPs) are binary systems that switch between a low-mass X-ray binary (LMXB) state
and a radio millisecond pulsar (MSP) one. 
These systems are very important since they are probes of the recycling formation scenario for MSPs, 
in which a neutron star (NS) is spun up by the accretion of mass from a 
low-mass non-degenerate companion \citep{1982-alpar}.

The binary PSR J1023+0038 was initially discovered by \citet{2002bond} as an optical source, 
but it was until \citeyear{2009archi} that it was discovered as a radio MSP \citep{2009archi}. 
Because of indications of recent accretion episodes in the system, previous to its detection as radio MSP, 
PSR J1023+0038 was the first system pointed out to be a tMSP \citep{2009archi}.
It has a spin period of 1.68 ms and an orbital period of 4.75 hrs.  
In mid-June of 2013 the object transitioned from a radio MSP to an LMXB state \citep{2014sta, 2014patruno,2014takata}, 
the current state of the binary. 

The system XSS J12270$-$4859 \citep{2004Sazo} first proposed as a tMSP by \citet{2011Hill}, has an orbital period of 6.91 hr \citep{2014bassa} and a spin period of 1.69 ms \citep{2015roy,2015papitto}. It was an accreting source up to November 14/December 21 of 2012, when the object decreased its brightness, switching to the MSP state \citep{2014bog,2014bassa}.
Up to date only one additional tMSP \citep[PSR J1824-2452I;][]{2013papitto} and one strong candidate \citep[1RXS J154439.4-112820;][]{2015candidate} have been identified.

Depending on their state (LMXB or MSP), tMSPs display interesting multi-wavelength properties. 
During the LMXB phase they are not visible as radio pulsar \citep[although they exhibit continuum radio emission; e.g.][]{2011Hill,2015deller}, 
but their emission in X-ray, gamma-ray and optical wavelengths 
increases in comparison to what they show in the MSP state \citep[e..g.][]{2013sta,2013martino,2014patruno,2014takata}. 
However, their X-ray luminosity is lower than that typically observed from LMXBs. 
No orbital modulation has been found in their X-ray light curves. X-ray pulsations are also observed in this state \citep[][]{2013papitto,2015papitto,2015Ar}.
Their optical spectra show double peaked emission lines, indicating the presence of an accretion disc \citep[e.g.][]{2014-martino,2014coti}. 

During the MSP state, the sources are detected as radio pulsars and radio eclipses are observed \citep[e.g.][]{2009archi,2014bog,2015roy}. 
Also, orbital modulation is seen in the X-rays and optical \citep{2014gentile}.  
It is thought that the X-ray modulation is produced by an intrabinary shock (created by the interaction of the pulsar wind and 
matter outflowing from the companion) near the companion star \citep{2011bog}. 
The optical modulations come from the combined effect of orbital variations and heating on the companion star (from the pulsar wind and/or intrabinary shock).
The result is an approximately single-humped light curve \citep[e.g.][]{2015-martino,2016-baglio}. 

The tMSPs have not been extensively studied in the ultraviolet (UV). 
In the MSP state, the UV is the transition region
where the dominant emission process might shift from the 
intrabinary shock (as seen in the X-rays) to the donor star (as seen in the optical).
Additionally, if another component in the system exists, such as a quiescent disc, it should manifest itself in the UV. 
Thus, to study better the intrabinary shock and/or to explore the existence of a quiescent disc, 
a UV study is necessary.
In this paper we focus on the first mid-UV (MUV) photometric observations of the tMSP XSS J12270$-$4859 during the MSP state.
Also, we present archival MUV images of PSR J1023+0038 during its MSP phase too.
Given that the published UV data of PSR J1023+0038 in that state is limited \citep{2014patruno,2014takata}, 
our results provide new insights about the behavior of the system in that phase.

\section{Observations and data reduction}

\subsection{XSS J12270$-$4859} 
\label{obs_XSS}

For the MUV analysis of XSS J12270$-$4859 during its radio MSP state, we obtained observations with the \textit{Hubble Space Telescope} (\textit{HST}) under program GO 13642 (PI: Degenaar). The observations were taken with the UVIS channel of the Wide Field Camera 3 (WFC3) on 2015 June 23 in a single visit. 
The data set consists of 16 images in the filters F218W (8 images), F225W (4 images) and F275W (4 images).
The total exposure times are given in Table \ref{foto}.
The native scale of the WFC3/UVIS images is 0.04 arcsec pixel$^{-1}$.
The source was detected in all the MUV images.
The photometric analysis of these images was carried out using the software \caps{DOLPHOT} \citep{dolphin}
which works on the individual .flc images. 
The individual images were barycentred and corrected for the charge transfer efficiency degradation\footnote{http://www.stsci.edu/hst/wfc3/ins$\_$performance/CTE/} of the UVIS detectors.
Photometric magnitudes were calibrated to the Vega system. 
We used a reddening value of \textit{E(B-V)}=0.11 \citep{2014-martino} towards XSS J12270$-$4859,
together with the reddening relation of \citeauthor{1989-cardelli} \citep[1989; including the update for the NUV given by][]{1994Odo} to account for 
extinction in each of the three \textit{HST} filters. The obtained photometric values are given in Table \ref{foto}. 

The photometric measurements of XSS J12270$-$4859 in the filters \textit{g', r'} and \textit{i'} used in this paper are those reported by \citet{2015-martino}.
They were obtained during the pulsar state from 2 sets of images taken with the ROSS2\footnote{http://www.rem.inaf.it} camera on the 0.6 m INAF REM telescope in La Silla, Chile. 
Their first set of images was obtained between 2015 January 20 and January 23, 
and their second one between 2015 February 12 and February 14. 

\begin{figure}
\centering
\vspace{-0.0cm}
  \includegraphics[width=5 cm, trim=1.5cm 0.0cm 0.0cm 1cm]{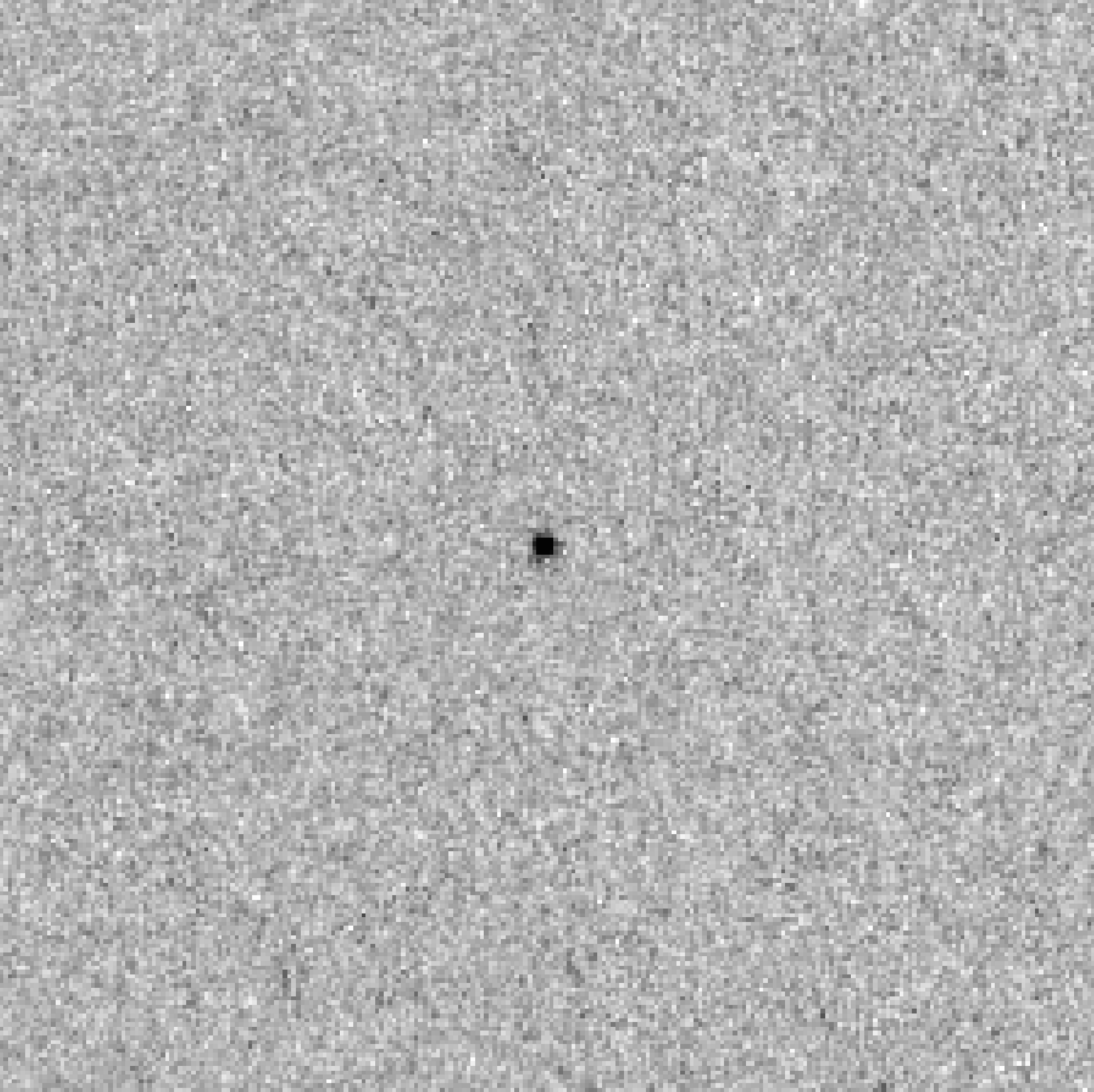} 
%\vspace{-1.3cm}
 \caption{\textit{HST}-F218W chart of XSS J12270$-$4859. The image is $10\arcsec \times 10 \arcsec$. }
\label{finding218}
\end{figure}

The photometric information in the bands \textit{J, H, K, i, R, V} and $B$ was obtained from \citet{2016-baglio}. 
Their images in $J, H$ and $K$ were taken on 2015 February 7 with SOFI\footnote{http://www.eso.org/sci/facilities/lasilla/instruments/sofi.html} 
on the ESO New Technology Telescope (NTT) at La Silla, Chile. Their images in $B, V, R$ and $i$ were obtained on 2015 February 8 and 9, 
with the EFOSC2\footnote{http://www.eso.org/sci/facilities/lasilla/instruments/efosc/inst.html} instrument also at NTT.
Additionally, we obtained and reduced short exposure ($\sim104$ s) $J, K$ and $H$ images taken with the Magellan telescope (Baade) on 2014 May 8. 
Photometry was performed using the photometric software \caps{IRAF/DAOPHOT} \citep{dao}. 
However, only photometric results in the $H$ band could be obtained. The $J$ and $K$ images were affected by bad pixels, inhibiting us to obtain reliable magnitudes. Therefore, these data are not used further in this paper.
We calibrated the $H$ magnitude to the Vega system using the 2MASS\footnote{http://www.ipac.caltech.edu/2mass/} catalogue (see Table \ref{foto}). 
The time resolution of the optical or MUV observations does not allow us to search for pulsations.

\subsection{PSR J1023+0038}
\label{psr_red}

\begin{table*}
\begin{center}
\begin{tabular}{ c | c |  c |   c |   c | c }
\multicolumn{6}{c}{Photometric results for XSS J12270$-$4859 and PSR J1023+0038 during their radio MSP states}\\
 \hline
&&XSS J12270$-$4859&&\\
\hline
Filter & Date & $\lambda_{eff}$& Mean  & A$_{\lambda_{eff}}$ & Total Exposure Time\\
&&(\AA)&Magnitude& (mag) & (s)\\
\hline
WFC3/F218W & 2015-06-23  & 2216 & $20.79\pm0.02$ &1.1 & 6436\\
''/F225W &''   &2359  & $21.82\pm0.03$ & 0.9&  2342\\
"/F275W &''  &2710  & $21.82\pm0.03$&0.7 &1980\\
\hline
ROSS2/$g'$	& 2015-01-20\textbackslash23 & 4640 & $18.62\pm0.02$  &0.42& 38916\\
"/$r'$	& '' & 6122 & $18.12\pm0.04$  &0.30&''\\
"/$i'$	& '' & 7439 & $17.96\pm0.02$ &0.22&''\\
\hline
"/$g'$	& 2015-02-12\textbackslash14 & 4640 & $18.75\pm0.03$  &0.42&55440\\
"/$r'$	& '' & 6122 & $18.28\pm0.03$  &0.30&''\\
"/$i'$	& '' & 7439 & $17.91\pm0.04$ &0.22&''\\
\hline
EFOSC2/$B$ 	&2015-02-08\textbackslash09  & 4345 & $18.95\pm0.13$  &0.45&480\\
"/$V$ 	& '' & 5458 & $18.32\pm0.05$  &0.34&''\\
"/$R$ 	& '' & 6416 & $17.95\pm0.05$  &0.28&''\\
"/$i$  	& '' &  7945 &$17.87\pm0.15$ & 0.21&''\\
SOFI/$J$ 	& 2015-02-07	& 12336 & $16.94\pm 0.06$ & 0.098&Unknown\\
"/$H$ 	& ''	& 16366 & $16.63\pm0.11$  &0.062&''\\
"/$K$	&''	& 21522 & $16.21\pm0.21$ & 0.040&''\\
\hline
$H$ 	& 2014-05-08	                       & 16366 & $16.65\pm0.16$  &0.062 &104 \\
\hline
&&PSR J1023+0038&&\\
\hline
UVOT-UVW2 (1) & 2013-06-10  & 2030 & $21.82\pm0.36$$^1$ & 0.66 &954 \\
UVOT-UVW2 (2) & ''  		       & ''        & $21.72\pm0.32$$^1$ & 0.66&967\\
UVOT-UVW1 (1) & 2013-06-12  & 2589 & $20.46\pm0.32$$^1$ & 0.49& 841\\
UVOT-UVW1 (2) & '' 		       & ''        & $19.54\pm0.14$$^1$ & 0.49 & 1024\\
\hline
\textit{GALEX}-NUV  (0)& 2009-03-07 & 2274 & $ >23^2$ & 0.67 &928\\
\textit{GALEX}-NUV  (1)& '' 			& '' & $19.96\pm0.08$ & 0.67 &1206\\
\textit{GALEX}-NUV  (2) & 2010-02-07 & '' &  $19.94\pm0.09$& 0.67 &768\\
\textit{GALEX}-NUV  (3) & '' 		& '' &  $20.48\pm0.12$& 0.67 &835 \\
\hline
\end{tabular}
\caption{Photometric results for XSS J12270$-$4859 and PSR J1023+0038. 
The effective wavelength ($\lambda_{eff}$) for each filter is given in column 3. 
Mean magnitudes for XSS J12270$-$4859  are given in the Vega system for the filters F218W, F225W, F275W, $B, V, R, J, H$ and $K$;  AB magnitudes are given for the filters $g', r', i'$ and $i$. For the system PSR J1023+0038 all the magnitudes are given in the Vega system. 
Uncertainties in magnitude correspond to $1\sigma$ errors.
The extinctions (A$_{\lambda_{eff}}$) in each filter were derived using the corresponding reddening value together with the relation of \citet{1989-cardelli}, including the update for the NUV given by \citet{1994Odo}. The parenthesis in the observations of  PSR J1023+0038 indicate the number of the image as shown in Figure \ref{finding}. }
{\flushleft
\startfoot  
\textfoot{Red leak corrections in the filters UVW1 and UVW2 have been taken into account and are 0.06 and 0.72 mags, respectively for a considered 6000 K black-body \citep{2010brown}.}
\textfoot{Derived using the faintest stars detected in the field of view of the image \textit{GALEX}-NUV (0).}
}
\label{foto}
\end{center}
\end{table*} 

We analysed MUV archival data of PSR J1023+0038 obtained with the near-UV filter of the Galaxy Evolution Explorer (\textit{GALEX}/NUV)\footnote{http://galex.stsci.edu/GR6/} satellite and with 
the Ultraviolet/Optical Telescope (UVOT) aboard \textit{Swift}\footnote{https://swift.gsfc.nasa.gov/archive/} while the object was in the radio pulsar state (i.e. before the transition in mid-June of 2013). 

The \textit{GALEX} data were taken in two visits, the first one on 2009 March 7 and the second one on 2010 February 7.
In total, four images were obtained (see Table \ref{foto}). We reduced the \textit{GALEX} data following the imaging analysis threads\footnote{http://www.galex.caltech.edu/researcher/techdoc-ch3.html} which uses
the software \caps{SExtractor} \citep{sectrac} to measure the fluxes of the target.  
Magnitudes were converted to the Vega system using the transformations given in \citet{luciana}.   

The \textit{Swift}/UVOT images were obtained on 2013 June 10 in the filter UVW2 (2 images) and on 2013 June 12 in the filter UVW1 (2 images). 
The total exposure times are given in Table \ref{foto}.
These images were reduced using the UVOT imaging analysis tools\footnote{http://www.swift.ac.uk/analysis/uvot/}.
We extracted magnitudes using circular regions of radius 3$\arcsec$ and $4\arcsec$ around the source for the UVW2 and UVW1 images, respectively. 
Unlike the other MUV filters used in this paper, the UVOT UVW1 and UVW2 filters suffer of red leak. 
Given the lack of a MUV spectrum of PSR J1023+0038 in the MSP state, 
we have taken the red leak corrections calculated by \citet{2010brown} for a 6000 K black-body (BB) spectrum.  
This temperature is consistent with that determined by \citet[][5650 K]{2005thor} for PSR J1023+0038.
We used the value of \textit{E(B-V)} = 0.073 \citep[determined by using the empirical relation
between the colour excess and the total column density of hydrogen,][]{2015-sha} and also the reddening curve of \citeauthor{1989-cardelli}
\citep[1989; including the update for the NUV given by][]{1994Odo}, to determine the extinction values in the \textit{GALEX} and UVOT filters. 
Details of the images and photometric results are given in Table \ref{foto}.

\begin{figure*}
\centering
\begin{tabular}{cc}
\large{Uniform priors with $0<q<200$ }\\
\vspace{0.1cm}
\twofigbox{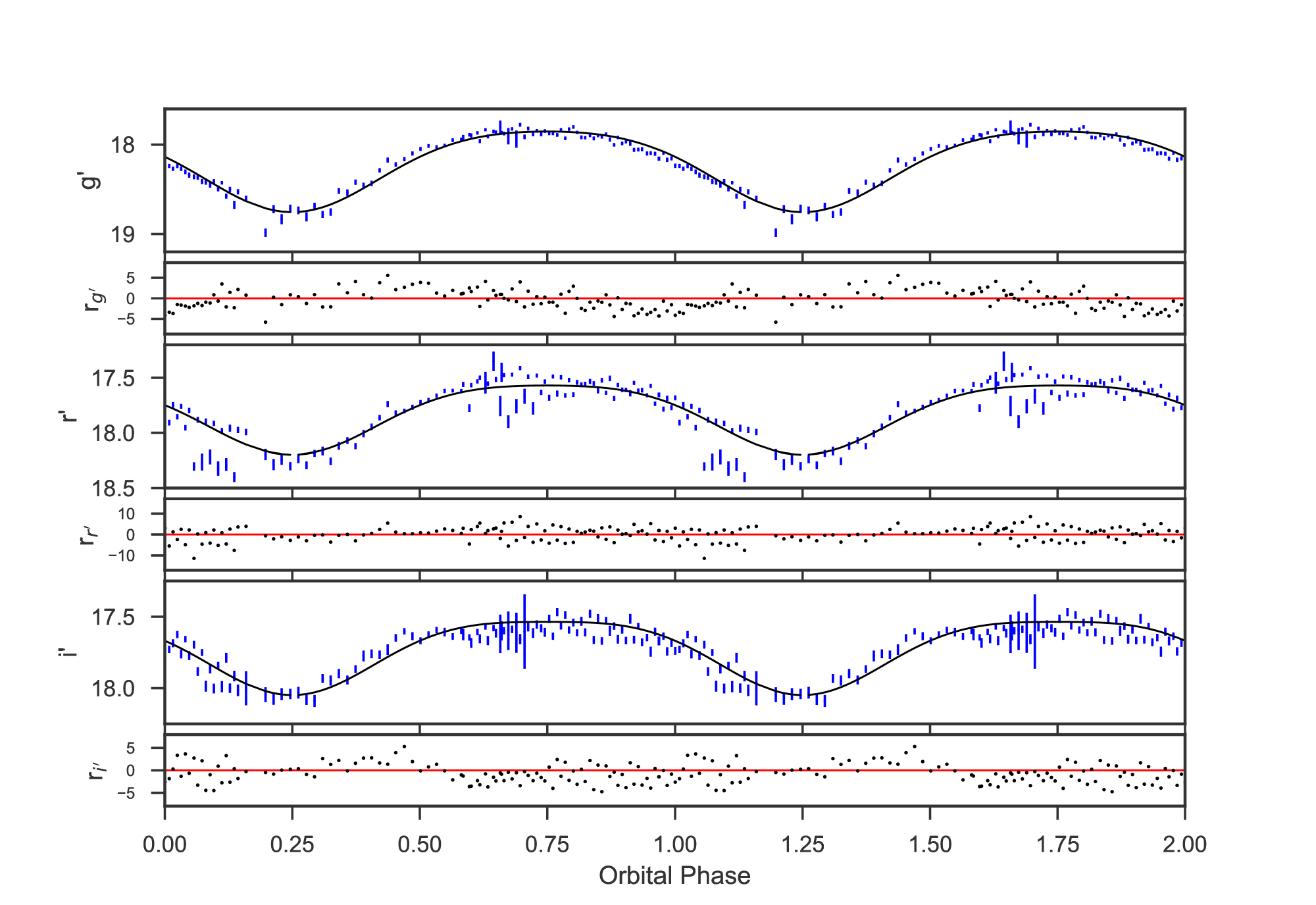}{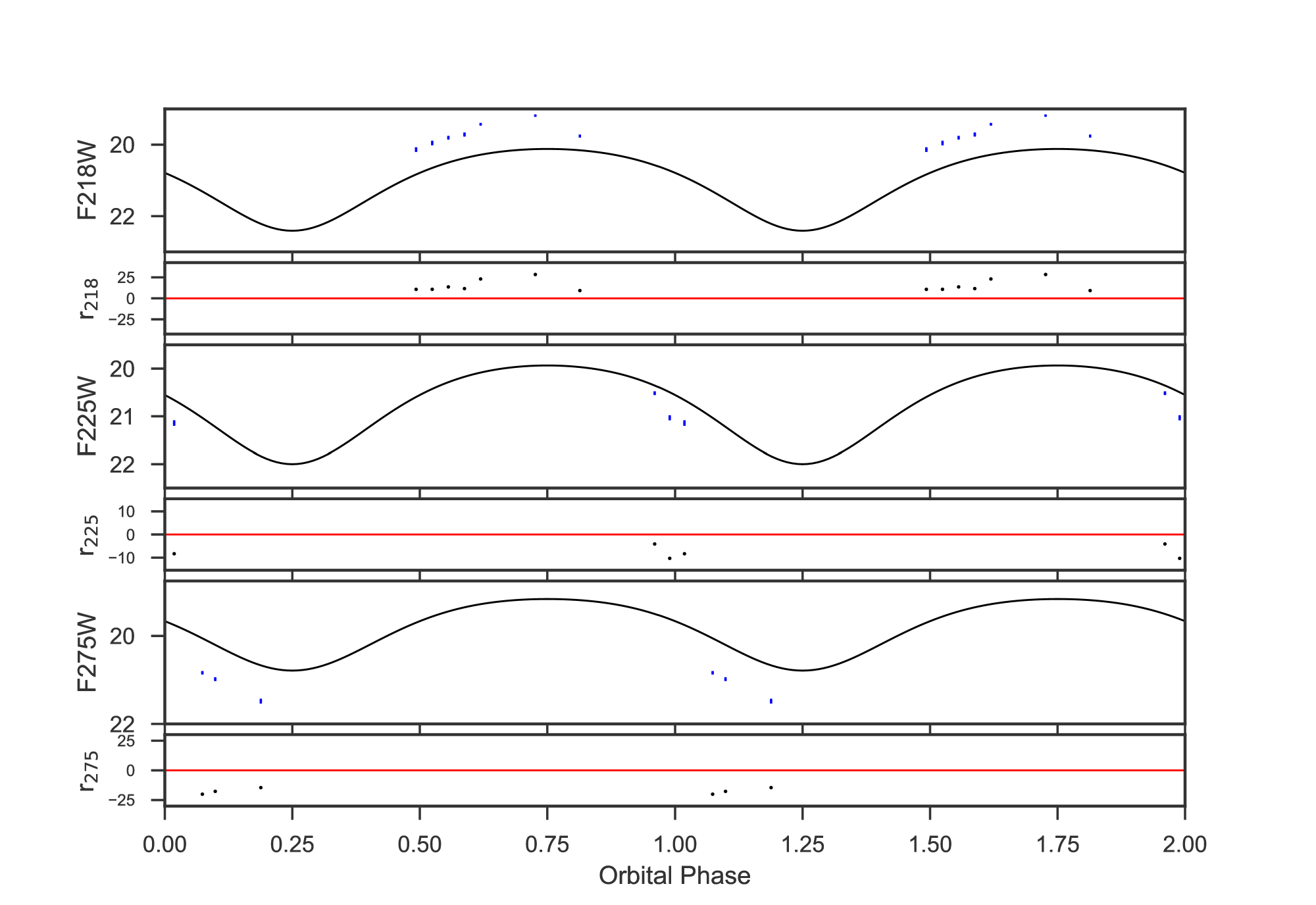}\\
\end{tabular}
 \caption{Phase folded light curves of XSS J12270$-$4859 in different UV filters (right) and in the filters $g', r'$ and $i'$ (left). For the latter we have used only the measurements from January 2015. The data of February 2015 present very large scatter (see Section \ref{parameters}). The best fit using MCMC is indicated with a solid gray line for the different priors of the considered parameters
(see Section \ref{parameters} for more information). 
Residuals (r) in each case are plotted below the corresponding light curves. 
These plots suggest the presence of systematic optical deviations in the system, which could affect the light curve fitting in these bands. 
Error bars represent the $1\sigma$ errors.
Phase 0.25 indicates that the companion is in inferior conjunction. 
Note how the modeled light curves in the filters $g', r'$ and $i'$ poorly fit to the data (${\chi}^2_{\nu}=7.3$). 
Using the same system parameters, the models completely fail to reproduce the MUV observations. 
A deficiency with respect to the models in the filters F225W and F275W is observed, whereas in F218W there is an excess. 
For clarity, the phase is plotted twice. 
The data were barycentred and dereddened before creating the folded light curves.}
\label{lightcurves}
\end{figure*}

\section{Results}

\subsection{System parameters of XSS J12270$-$4859}
\label{parameters}

XSS J12270$-$4859 was detected in all our MUV images.
There were not stars nearby that could contaminate the photometry (Figure \ref{finding218}).  
However, three measurements (one per filter) were discarded because they were affected by bad pixels.  
We only used the remaining good measurements to build light curves in the different MUV filters. 
The source was clearly variable in all MUV filters. To determine if this variability was related to the orbital motion, we used the
orbital period and time of ascending node of the pulsar given by \citet[][ephemeris obtained with $Parkes$ observations]{2015papitto} to phase fold the MUV light curves of XSS J12270$-$4859 (Figure \ref{lightcurves}, right panels). The orbital phase 0.25 corresponds to the inferior conjunction of the donor star (i.e. when it is in between the observer and the NS).
From these folded curves we see that the system shows large variations in all the MUV filters. 
In the filter F218W, XSS J12270$-$4859 shows an amplitude variation of $\sim 0.9$ mags between phases $\sim0.5$ and $\sim0.75$. 
The data in the filters F225W and F275W show amplitude variations of $\sim0.6$ mags in the covered phase ranges ($\lesssim 0.1$). 

In order to determine the mass ratio of the system, the temperature of the companion, the orbital inclination and the distance to XSS J12270$-$4859, and compare them to those of previous studies \citep[e.g.][]{2014-martino, 2015-martino, 2016-baglio}, we carried out a folded light curve fitting using the data in the filters \textit{g', r'} and \textit{i'} (magnitudes and errors kindly provided by D. de Martino).
These observations are simultaneous, were taken only few months before our MUV observations, and cover the entire orbital period of the system \citep[see][for more information]{2015-martino}. To fold the optical observations we again used the ephemeris of XSS J12270$-$4859 given in \citet{2015papitto}. 

For our fitting we used the stellar binary light curve synthesis tool \textsc{icarus}\footnote{https://github.com/bretonr/Icarus} \citep{2012Breton}.
This code calculates the incoming flux from a star at each orbital phase and inclination. 
We modeled the companion star using the stellar atmosphere grids CIFIST2011$\_$2015 \citep{2015-bara}
in the corresponding filters.
These atmospheres are suitable for stars in the effective temperature range $1200 <T_{eff}<7000$ K. 
That range covers the expected temperatures for the companion in XSS J12270$-$4859. 
We implemented the \caps{icarus} tool and the stellar atmosphere models in a Markov Chain Monte Carlo (MCMC) algorithm \citep[\caps{emcee};][]{2013mcmc} to determine the following parameters:

\begin{itemize}
\item $q$ - Mass ratio M$_{NS}/$M$_{C}$, where M$_{C}$ is the mass of the companion star and M$_{NS}$ the mass of the NS. 
\item $T_{base}$ - Surface temperature of the star before irradiation and gravity darkening. 
\item $T_{irr}$ - Irradiation temperature, with the irradiation coming from the NS (e.g., due to the pulsar wind) 
\item i - Inclination of the system
\item $\mu$ - Distance modulus to XSS J12270$-$4859.   
\end{itemize}

In our calculations the effects of gravity darkening \citep[$\beta=0.08$,][]{1967lucy} and limb darkening \citep{2011claret} were taken into account. We have also assumed that the system is tidally locked and 
that the donor fills its Roche lobe. Results for smaller filling factors (0.8, based on \citeauthor{2015Mac} \citeyear{2015Mac} for PSR J1023+0038)
are consistent (within errors) with those of complete Roche lobe filling. 

\begin{figure}
\centering
\vspace{-0.3cm}
  \includegraphics[width=9.5cm, trim=1.5cm 0.0cm 0.0cm 1cm]{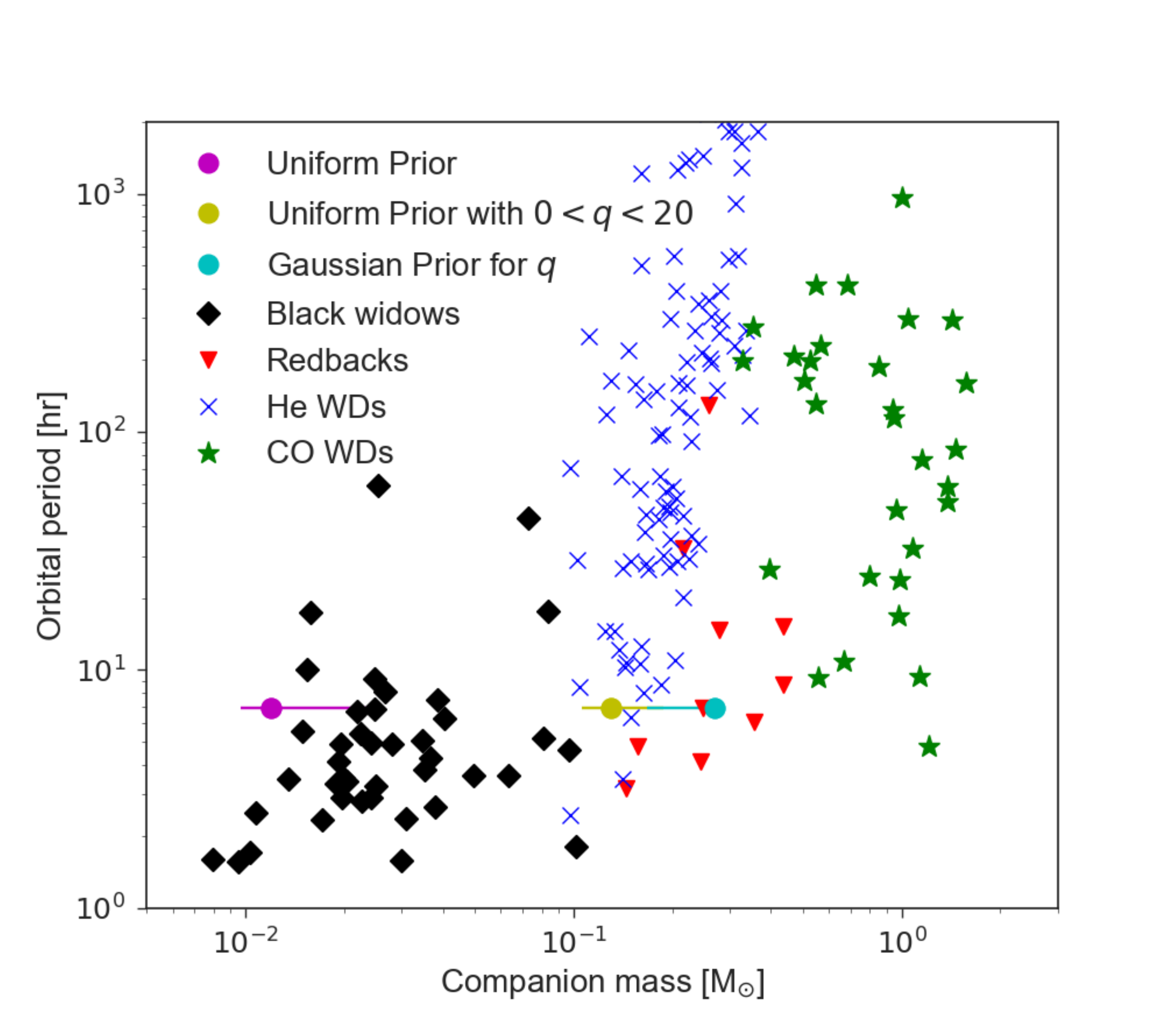} 
%\vspace{-1.3cm}
\vspace{-.6cm}
 \caption{For different types of known pulsars, the median mass of the companion versus orbital period of the pulsar are plotted.
Data taken from the \textit{ATNF Pulsar Catalogue}$^{11}$ in June 2017, 
 \citep{2005manches}. The mass ranges for the companion obtained from our three different fits are also shown (horizontal lines).}
\vspace{-0.5cm}
\label{black}
\end{figure}

We individually fitted the data from January 2015 and the data from February 2015. 
This is because the data from February presents very large scatter, specially in the bands $r'$ and $i'$, which significantly affects the fitting results. 
However, the values of $T_{base}$, $T_{irr}$, i and $\mu$ obtained from February are in agreement with those of January. 
The value of $q$ was several times substantially affected by the large magnitude scatter in the data from February. 
For this reason in the following analysis we focus only on the data from January. 
We first fixed the mass ratio and distance to XSS J12270$-$4859 (thus the distance modulus) leaving the temperatures and
inclination free to vary. 
For this, we used the value $q=5.15(8)$ given by \citet{2015roy}, which was determined based on their radio measurement of the projected radial velocity amplitude of the pulsar and the spectroscopic measurement of the projected radial velocity amplitude of the companion 
\citep[][]{2014-martino}. Such value of $q$ is also in the range given by \citet{2015-martino}. 
We also used $d=1.4$ kpc \citep[as obtained from radio measurements;][]{2015roy}.
With these assumptions the orbital inclination did not converge. 
Thus, to reach convergence in all the variables (i.e. that the Markov chains appear to be sampling from regions with high probability) we left them free to vary. 
In our MCMC algorithm we defined uniform priors for all the five variables, we used 2000 walkers and 700 steps per walker. 
We imposed a range of values for the parameter $q$ between 0 and 200.
This upper limit is based on the lowest median masses of some black widow companions (see Figure \ref{black}) and it is
equivalent to 0.007 M$_{\sun}$ for a NS of 1.4 M$_{\sun}$. 
Under these assumptions all the parameters reached convergence. 
The results obtained are shown in Table \ref{mcmc_par} and in Figure \ref{uniform} of the Appendix \ref{apen1}.

We noted that the probability distribution of $q$ shows an smaller probability peak at around $q=10$ (Figure \ref{uniform}). 
Therefore, to explore that range of values we reduced the domain of $q$ from (0,200) to (0,20). 
Again, we assumed uniform priors for the five parameters. We used now 2000 walkers and 500 steps. 
Results are also presented in Table \ref{mcmc_par} and 
shown in Figure \ref{posteriors} of the Appendix \ref{apen1}, where we observe that indeed there is a maximum of probability for $q$ at around 11
which corresponds to M$_C=0.13$ M$_{\sun}$. 

\footnotetext[11]{http://www.atnf.csiro.au/people/pulsar/psrcat/}

To take into account in our MCMC algorithm the mass ratio given by \citet[][$q=5.15(8)$]{2015roy},  
we defined a Gaussian prior for the parameter $q$, 
leaving uniform priors for the other four parameters.
All the parameters converged as well using that approach (Figure  \ref{posteriors_gauss} of the appendix). 
All our results are discussed more in detail in Section \ref{dis}.

To investigate the effect of reddening, we also tried to add \textit{E(B-V)} as another free parameter. 
However, the solution for this parameter did not converge despite that
the parameters $q$, $T_{base}$, $T_{irr}$, i and $\mu$ converged. 
In fact, the obtained median values for these five parameters are consistent (within the error limits) with the values obtained using 
these parameters alone (i.e. without including \textit{E(B-V)}). 
Therefore, to investigate the effect of \textit{E(B-V)}, we manually set this parameter to different values (0.05, 0.2) and re-run the MCMC algorithm with five parameters. All of them converged under that approach. The results for $q$, i and $\mu$ are similar for all the values of \textit{E(B-V)}. 
However, we noted that the resultant temperatures were correlated to the changes in \textit{E(B-V)}. 
For a larger value of  \textit{E(B-V)}, we obtained a larger value of $T_{base}$ and $T_{irr}$.
This is understandable since for a given observed magnitude (or flux) 
the star has to be hotter to compensate for the loss of flux due to the higher reddening (thus extinction) values.

\begin{table}
\caption{System parameters obtained for the tMSP XSS J12270$-$4859 using our folded light curve fitting method.
The errors in our values represent the 16\% and 84\% percentiles, which would correspond to 1$\sigma$
errors in a Gaussian distribution.
Since the fitting is not optimal, the values presented below should be taken as rough indicators and limits.   
For comparison, value ranges at 1$\sigma$ reported in the literature are also given.}
\label{mcmc_par}
\begin{center}
\begin{tabular}{ l | c | c | r }
\hline
Parameter & Median &  $16-84$& Literature\\
&&percentile&\\
 \hline
\multicolumn{4}{c}{Uniform priors}\\
\hline
$q$				 	& 110.9 &52-136& 3.8-9.1$^1$, 5.1-5.2$^2$\\
M$_{C}$ (M$_{\sun}$) 	& 0.012   &0.010-0.024 &0.15-0.36$^1$\\
i ($^{\circ}$) 		         &  34       & 23-47&46-65$^1$ \\
$\mu$ 				&10.3   &10.1-11.9& 10.73 \\
Distance (kpc) 			&1.17&1.07 -2.44 &1.4$^2$, 1.24$^3$\\
$T_{base}$ (K) 			&5351&5305-5617& - \\
$T_{irr}$ (K) 			&6609&6007-6671& - \\
$\langle T_{day} \rangle$  (K) &6357&-& 6500$^1$\\
$\langle T_{night} \rangle$  (K) &5197&- &5500$^1$\\
\hline
\multicolumn{4}{c}{Uniform priors with $0<q<20$}\\
\hline
$q$				 	& 10.6 &7.3-13.0& 3.8-9.1$^1$, 5.1-5.2$^2$ \\
M$_{C}$ (M$_{\sun}$) 	& 0.13 & 0.10-0.19 &0.15-0.36$^1$\\
i ($^{\circ}$) 		         &  38   & 32-49 &46-65$^1$ \\
$\mu$ 				& 11.9 &11.7-12.9& 10.73 \\
Distance (kpc) 			&  2.39 &2.18-3.80  &1.4$^2$, 1.24$^3$\\
$T_{base}$ (K) 			& 5457 & 5510-5615& - \\
$T_{irr}$ (K) 			& 6216 &5944-6307& - \\
$\langle T_{day} \rangle$  (K) &6379&-& 6500$^1$\\
$\langle T_{night} \rangle$  (K) &5324&- &5500$^1$\\
\hline
\multicolumn{4}{c}{Gaussian prior for $q$}\\
\hline
$q$					& 5.2 & 5.1-8.4 & 3.8-9.1$^1$, 5.1-5.2$^2$ \\
M$_{C}$ (M$_{\sun}$) 	& 0.27 & 0.17-0.27&0.15-0.36$^1$\\
i ($^{\circ}$) 		         &  40 & 34-46&46-65$^1$ \\
$\mu$ 				& 12.5 & 12.0-12.6& 10.73 \\
Distance (kpc) 			& 3.16 & 2.51-3.31 &1.4$^2$, 1.24$^3$\\
$T_{base}$ (K) 			& 5501 & 5380-5544   & - \\
$T_{irr}$ (K) 			& 6037 & 5977-6231  & - \\
$\langle T_{day} \rangle$  (K) & 6385 &-& 6500$^1$\\
$\langle T_{night} \rangle$  (K) & 5375 &- &5500$^1$\\
\hline
\end{tabular}
{\flushleft
\startfoot  
\textfoot{Taken from \citet{2015-martino}, M$_{C}$ has been derived considering $M_{NS}=1.4$ M$_{\sun}$.}
\textfoot{Taken from \citet{2015roy}.}
\textfoot{Using the YMW16 model, \citet{2017yao}.}
}
\end{center}
\end{table}

Once we had the best estimated values from the MCMC runs 
we integrated over the corresponding surfaces of the companion star to get the average temperatures 
of the irradiated (facing the NS, $\langle T_{day} \rangle$) and the non-irradiated side ($\langle T_{night} \rangle$, see \citeauthor{2016her}, \citeyear{2016her} for a detailed description of the temperature formulae). Results are given in Table \ref{mcmc_par}.

We have used the obtained MCMC parameters to over plot the corresponding model in
each folded light curve in Figures \ref{lightcurves} and \ref{lightcurvesappen} of appendix \ref{apen0}. 
As described before, these parameters were derived using only the optical data. 
We note that for the folded light curves in the filters $g', r'$ and $i'$ the model poorly describes the observations (${\chi}^2_{\nu}=7.3$ in all cases). 
This could be partially due to possible systematic deviations in the optical data, as suggested by the residual plots (Figures \ref{lightcurves} and \ref{lightcurvesappen}, left panels).
Thus, to test the effect of this in our MCMC solution, we have artificially increased their value by a factor of $\sim3$, 
ensuring in this way a ${\chi}^2_{\nu}=1$. 
The MCMC solutions were very similar to the ones presented in Table \ref{mcmc_par}. 
This happens because the relative error between points is the same, as well as their weight in the fit. 
Therefore, the values presented in Table \ref{mcmc_par} can be considered as limits for a direct heating model. 

When we use the same parameters to model the \textit{HST} MUV light curves, the models completely fail to reproduce our MUV observations (differences between systems AB and Vega are taken into account in our fittings).
Deficiencies were detected for the filters F225W and F275W using uniform or Gaussian priors for the mass ratio (see Figures \ref{lightcurves} and \ref{lightcurvesappen}, right panels).  
In all the cases, the observations in the filter F218W showed an excess with respect to the model.
The poor fitting in optical and MUV, which results from assuming only direct heating on the companion,
shows that a more complete model (with one or multiple extra components) should be taken into account to reproduce the observations of XSS J12270$-$4859 in these bands. 

\subsection{Spectral Energy Distribution of XSS J12270$-$4859}
\label{sed}

\begin{figure}
\centering
\vspace{-0.3cm}
  \includegraphics[width=9cm,trim=0.0cm 0.0cm 0.0cm 1.5cm]{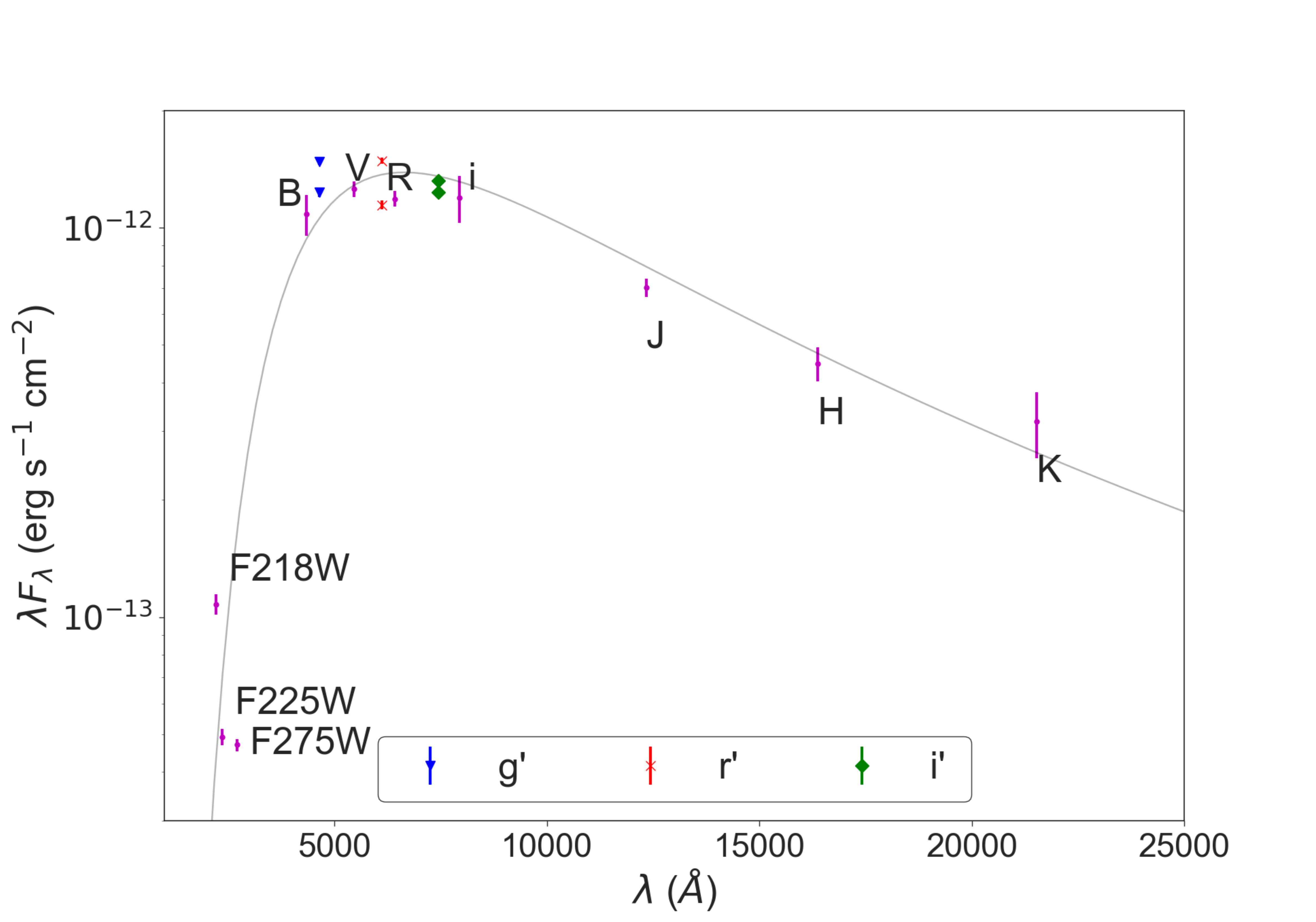} 
 \caption{SED of XSS J12270$-$4859 including NIR, optical and MUV observations. 
The solid line represents the best fit using a black-body model with temperature $T_{BB}=5508 \pm 159$ K. For clarity,  the filter is indicated for each point. 
The fluxes in the filters F218W, F225W, $g', r'$ and $i'$ correspond to an orbital phase of 0.52. The flux in the filter F275W is given for an orbital phase of 0.42 and was not taken into account for the fitting (see Section \ref{sed}). The data points in the filters $B, V, R, i, J, H$ and $K$ correspond to an average orbital phase. All the fluxes have been corrected for extinction. Note that the model fails to fit the observations in the filters F218W, $B$, $g'$ and $J$. }
\label{bbsed}
\end{figure}

Combining the available optical and near-infrared (NIR) data of XSS J12270$-$4859 \citep{2015-martino,2016-baglio}
with our MUV results, we constructed a broad band spectral energy distribution (SED). 
All the flux densities were corrected by the corresponding extinction value. 
Given the large amplitude variations found in MUV, and the incomplete sampling of the orbit, 
we built the SED for approximately the same orbital phase. 
To do this, we assumed that the real curves (and therefore the current observations) in the MUV filters F225W and F275W are symmetric. 
This assumption is based on the symmetric folded light curve obtained by \citet{2015-martino} in the U band ($\lambda_{eff}=3440$ \AA)
using the optical monitor aboard the X-ray satellite \textit{XMM-Newton} (OM-U).
Under the assumption of symmetry, our observations in F225W (and approximately also in F275W) cover similar phases as those in F218W.  
Therefore we chose to build the SED at an orbital phase of $\sim0.52$, 
which is covered by the observations in the filters F218W, F225W, $g', r'$ and $i'$. 
Exceptions are the observations in the filters F275W, for which the closest phase is 0.42, and $B, V, R, i, J, H, K$ for which we only have average magnitudes. 
However, since the amplitude variations get smaller as we move towards the NIR \citep[see light curves in][]{2016-baglio},  
taking the mean magnitudes for these observations is a reasonable approach. 
The resultant SED is shown in Figure \ref{bbsed}. 
For consistency with the dates of observation, we have used the $H$ magnitude from \citet{2016-baglio} in our SED. 
However, we note that our derived magnitude in the $H$ band (image taken $\sim9$ months earlier) is consistent with the one derived by those authors. 
This indicates that the emission in that band has been approximately constant for XSS J12270$-$4859 during the pulsar state. 

We fitted a BB model to the SED and we obtained a temperature $T_{BB}=5508 \pm 159$ K, 
which is consistent with the one found by \citet{2015-martino}. 
For this fit we have included the observations in $g', r'$ and $i'$ from February (Section \ref{obs_XSS}), but we excluded the MUV observation in the filter F275W since it corresponds to a different orbital phase (0.42) than F218W and F225W. 
This small difference in phase affects substantially the flux in that band. Note that our observations in that filter 
cover a phase range of $\sim0.1$ and show amplitude variations of $\sim0.6$ mags. 
These kind of large amplitude variations could account for the apparent flux deficiency in F275W with respect to the BB fit. 
Note also how the BB model fails to reproduce the observations in the bands F218W, $B$, $g'$ and $J$.  
The deficiencies and excess in the different MUV filters seem to indicate that we are seeing different components of the system at different wavelengths. 
The companion star largely contributes to the optical and likely dominates the NIR emission, but the MUV can not come fully from it (an excess of approximately half an order of magnitude in F218W is observed). To investigate this, we have fitted a BB model only to the NIR data in the bands $J, H$ and $K$ assuming that indeed it is the companion star which dominates the emission at these wavelengths. We obtained a fit ($T_{BB}= 4350$ K) that substantially underestimates all the optical and MUV observations.
This does not imply that the companion is not largely contributing to the optical flux (the NIR observations do not cover a very large wavelength range, which biases the NIR fit). Instead, it suggests that the extra component could also significantly contribute to the observed optical flux, and also that its contribution becomes more important in the MUV. Therefore a simple BB model fit would fail reproducing the data from the MUV to the NIR bands, as already observed from Figure \ref{bbsed}. 
This means that our derived value of $T_{BB}=5508\pm159$ K should be taken as an upper limit.

\subsection{UV variations of PSR J1023+0038 in the MSP state}
\label{psr}

In order to understand better the configuration of the system PSR J1023+0038 during its MSP state, 
we analyzed MUV images taken with \textit{GALEX}, we also re-analyzed those images taken with $Swift$/UVOT. 
We have used the ephemeris given by \citet{2016jao} to phase fold the light curve of the PSR J1023+0038 on the orbital period. We have corrected the data for extinction, red leaks (in the case of UVOT filters) and we have barycentred the $Swift$/UVOT and \textit{GALEX} measurements. The folded light curves of PSR J1023+0038 in different UV filters in the MSP state are shown in Figure \ref{GALEX_lc}. Charts of the object at different orbital phases are given in Figure \ref{finding}.
From these images we see that PSR J1023+0038 was not detected (m$_{MUV} >$ 23 mag) in the MUV by \textit{GALEX} in 2009, at an orbital phase of $\sim0.09$ (at phase 0.25 the companion star is in inferior conjunction). This contrasts with the observation of \textit{GALEX}/NUV in 2010 at an orbital phase of 0.93, in which the object was clearly detected with a shorter exposure image (928 s vs 835 s). Another interesting result is that the system was not detected at phase $\sim0.09$ (m$_{MUV} >$ 23 mag) in the 
\textit{GALEX}/MUV observations of 2009, but few hours later the system was detected at an orbital phase of 0.43 (near the passage of the descending node) with m$_{MUV} =$ 19.96 mag. In Section \ref{dis} we discuss more in detail the implications of such behavior. 

Regarding the UVOT images, the source was detected in the UVW2 filter on June 10 of 2013 (see Figure \ref{finding}), contrary to the findings of \citet{2014takata} and \citet[][who only provided an upper limit of 20.6 mag in the UVW2 filter]{2014patruno}. In fact, we detected it at orbital phases of 0.43 and 0.82. Similarly, the object was detected with the filter UVW1 on June 12 of 2013 \citep[few days before the transition of the system from the MSP state into the LMXB one on $\sim15$ June 2013;][]{2014sta} at orbital phases 0.55 (brightest measurement among the UVOT data points) and 0.43. Our UVOT photometry performed on the stacked images and corrected by red leaks, indicates that the system was detected with average reddened magnitudes of $21.72\pm0.31$ in UVW2 and  $19.92\pm0.17$ in UVW1. Our average magnitude in UVW1 is consistent with the UVW1 magnitude given by \citet{2014patruno}.

\section{Discussion}
\label{dis}

We have reported on the first MUV data, obtained with \textit{HST}, of the tMSP XSS J12270$-$4859 in its MSP state. 
We clearly detected the source and found it variable during the time of our observations. In addition, we have also reported on \textit{GALEX} and \textit{Swift}/UVOT data of the tMSP PSR J1023+0038 taken during the radio pulsar state. This source was also detected in MUV and it was found to be strongly variable. In the following section we discuss the implications of our results.

\subsection{Implications for the companion's mass and distance to XSS J12270$-$4859 under the direct heating model}

\begin{figure}
\centering
\vspace{-2.3cm}
  \includegraphics[width=8.5 cm, trim=1.5cm 0.0cm 0.0cm 0cm]{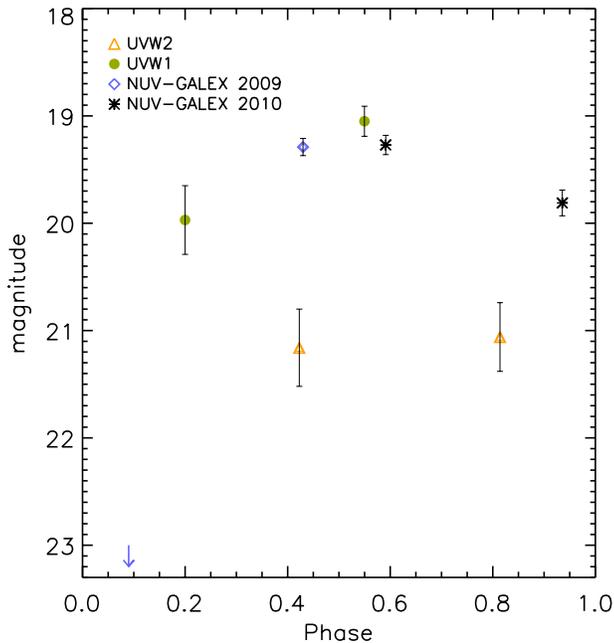} 
\vspace{-1.3cm}
 \caption{Phase folded light curve for the $Swift/$UVOT and \textit{GALEX} measurements of PSR J1023+0038. The data have been dereddened and barycentred. Phase 0.25 indicates that the companion is in inferior conjunction. The upper magnitude limit ($m_{MUV}>23$) for the image \textit{GALEX}-NUV (0) is indicated with a purple arrow. All the data have been taken during the radio MSP state.}
\label{GALEX_lc}
\end{figure}

In Sections \ref{parameters} and appendix \ref{apen1}, we show the results of our joint posterior probability distributions for uniform priors using MCMC.
We note that the mass ratio displays a global maximum at $q=110.9$, which corresponds to a companion's mass of $\sim0.012$ M$_{\sun}$. This is consistent with median masses of some black widow\footnote[12]{Black widows and redbacks are companions to radio MSPs that are being ablated by the pulsar.  
Black widows have masses M$_C\lesssim0.1$ M$_{\sun}$, whereas redbacks can have masses up to $\sim0.4$ M$_{\sun}$.} systems (see Figure \ref{black}), such as pulsar PSR J1311-3430 which has a companion with mass $0.01$ M$_{\sun}$ \citep{2015romani}.
A low mass companion is also derived for the local maximum at $q=10.6$. 
In this case the companion's mass ($0.13$ M$_{\sun}$) is an intermediate value between the observed median masses of black widow and redback systems. 
This means that if the real mass of XSS J12270$-$4859 is indeed $\lesssim0.13$ M$_{\sun}$, the system could be a black widow (or an intermediate system, see Figure \ref{black}) instead of a redback, as believed so far. In such case, it is possible that XSS J12270$-$4859 harbors a very massive NS, according to the evolutionary scenario of \citet[][]{2014ben} for black widows. In fact, there is some observational evidence in other black widows which points towards the existence of massive NS \citep{2011van,2012romani}.
Note that in this case our derived mass is in agreement (within errors) with the value obtained from spectroscopic observations by \citet{2014-martino}, who identified the companion as a resembling mid G-type star with mass $0.06-0.12$ M$_{\sun}$. Similar low masses and (resembling) spectral types have been found for other binary pulsars, such as the black widows system PSR B1957+20 which companion has been classified as a G-type star with a mass of $\sim0.035$ M$_{\sun}$ \citep{2011van}.
Our value of M$_C$ = 0.27 M$_{\sun}$ obtained using a Gaussian prior is not informative, since such value 
was assumed to define the prior of the parameter $q$, thus it was expected that the best fit value would be close to that number. Given the problems that the direct heating model has to explain our current observations, none of our $q$ values is preferred over the others.

From our fit we also note that the estimated distance towards XSS J12270$-$4859 is 
approximately in agreement with the distance determined from radio dispersion measure (DM) if the mass
of the companion is extremely low ($\sim0.012$ M$_{\sun}$).
In the other cases, the distance is $\sim1$ kpc and $\sim2$ kpc
larger than that of DM \citep[1.4 kpc,][]{2015roy}. 
That discrepancy depends on our assumptions of the priors for the mass ratio. 
The largest difference appears when we take into account a Gaussian prior for $q$, i.e. when we consider the
information obtained from radio and optical spectroscopic measurements. 
However, two of our median values of the distance (2.4 and 3.2 kpc) are within the range determined by \citet{2013martino}. These authors used UV/optical and NIR spectra of XSS J12270$-$4859 during the LMXB state and derived a distance of 2.4-3.6 kpc. The discrepancies between the distances obtained from our light curve fitting using MCMC and that from DM, seem to suggest that XSS J12270$-$4859 has a 
companion with a mass rather small (assuming a NS mass of $1.4$ M$_{\sun}$).   
However, recently \citet{2017romani} have shown that optical light curve fittings that only assume 
direct heating by the pulsar wind of the NS on the companion (as we have assumed in this work), usually require distances that are larger than those determined by DM. 
They propose a heating model in which an intrabinary shock reprocesses the pulsar wind. 
This reprocessing produces particles that are ducted to the companion's magnetic
poles, producing hot spots on it. That mechanism would provide additional heating, helping to solve the problem of distances and luminosities observed. 
This means that the shock could also be contributing indirectly in the optical band and not just in the UV. 
These authors also found that if the distances are assumed to be the ones determined from radio DM, 
their model provides better light curve fittings than only considering direct heating. 
Furthermore, the masses of the components would have more reasonable values compared to the case of direct heating. 

\begin{figure*}
\centering
  \begin{tabular}{@{}cccc@{}}
   \includegraphics[width=.23\textwidth]{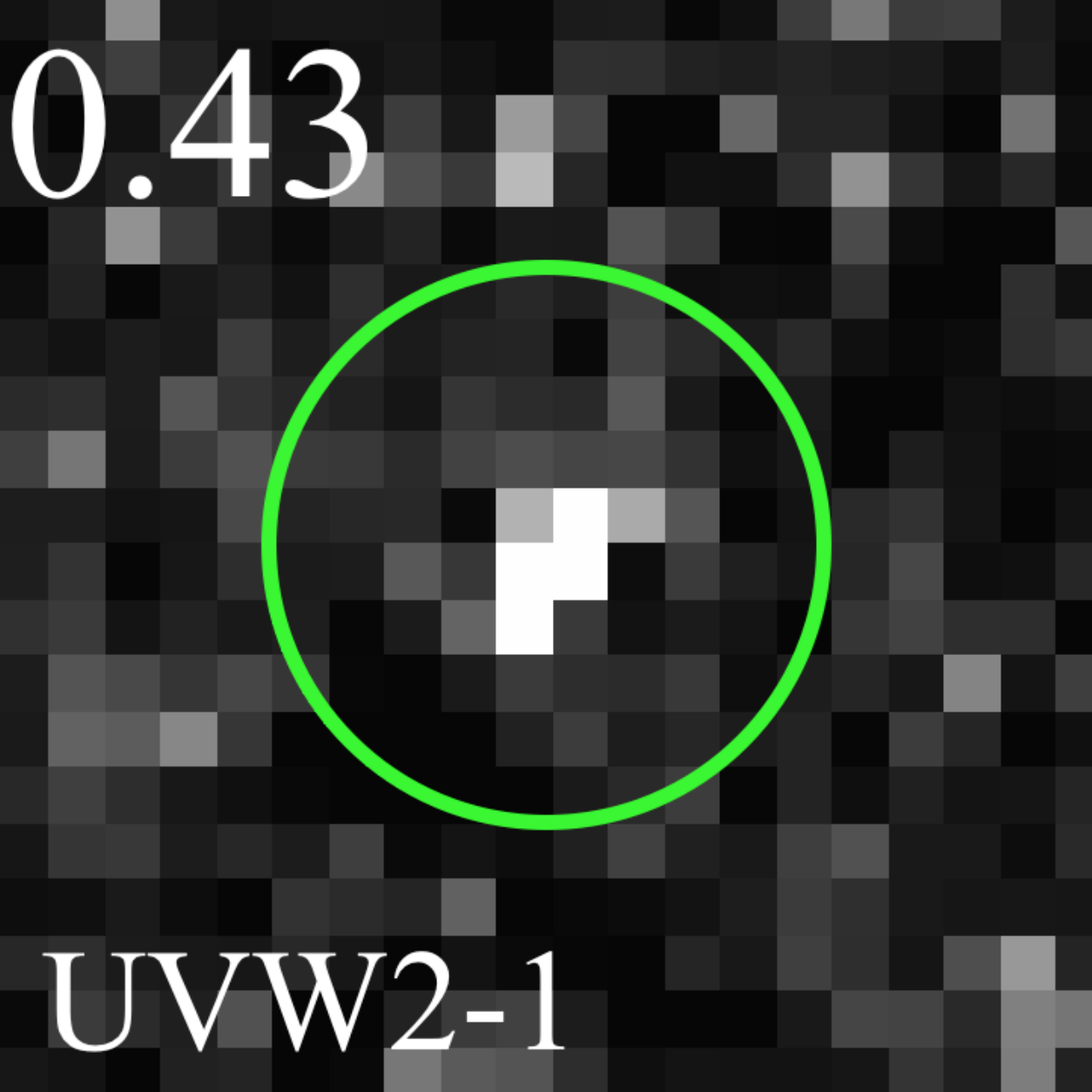} &
   \includegraphics[width=.23\textwidth]{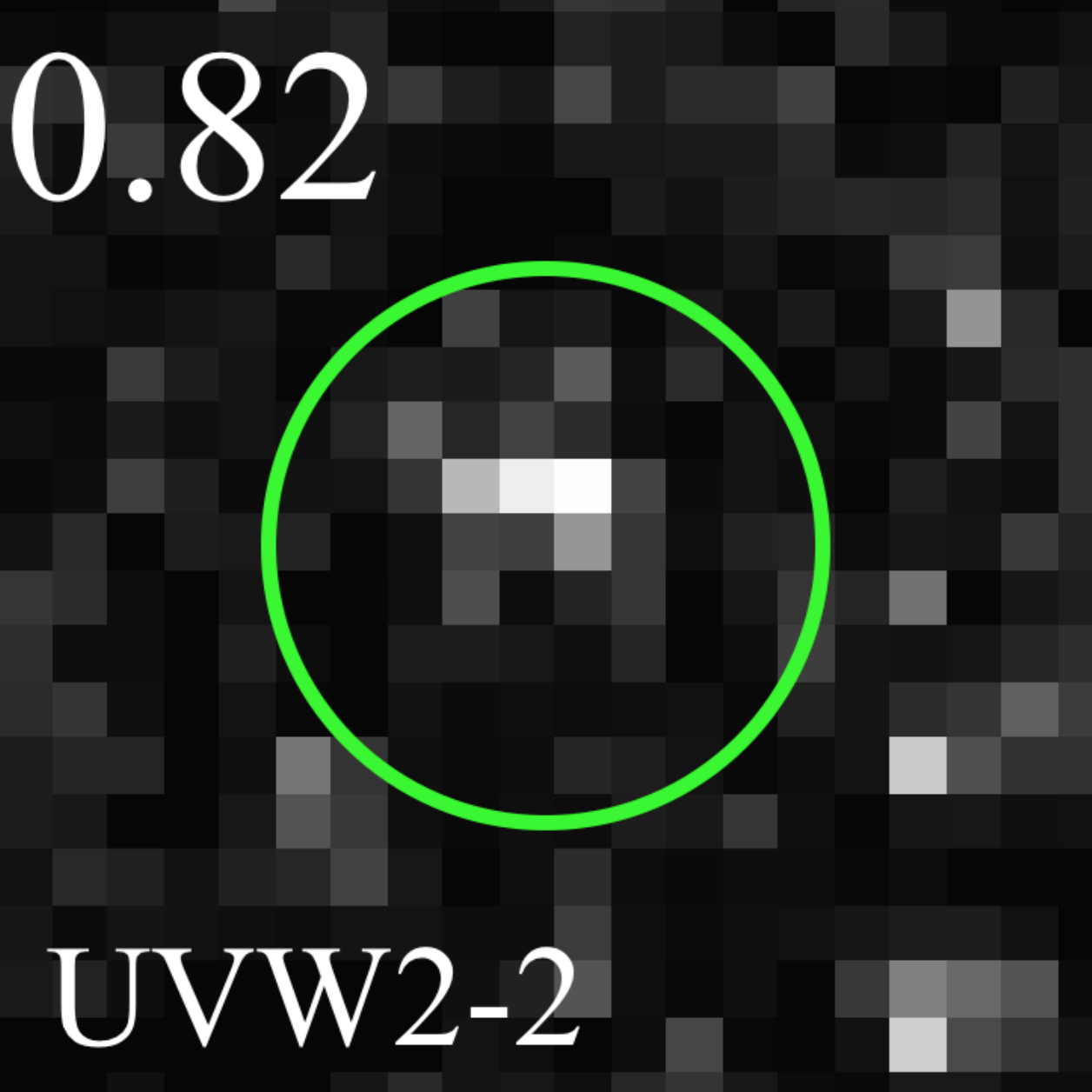} &
  \includegraphics[width=.23\textwidth]{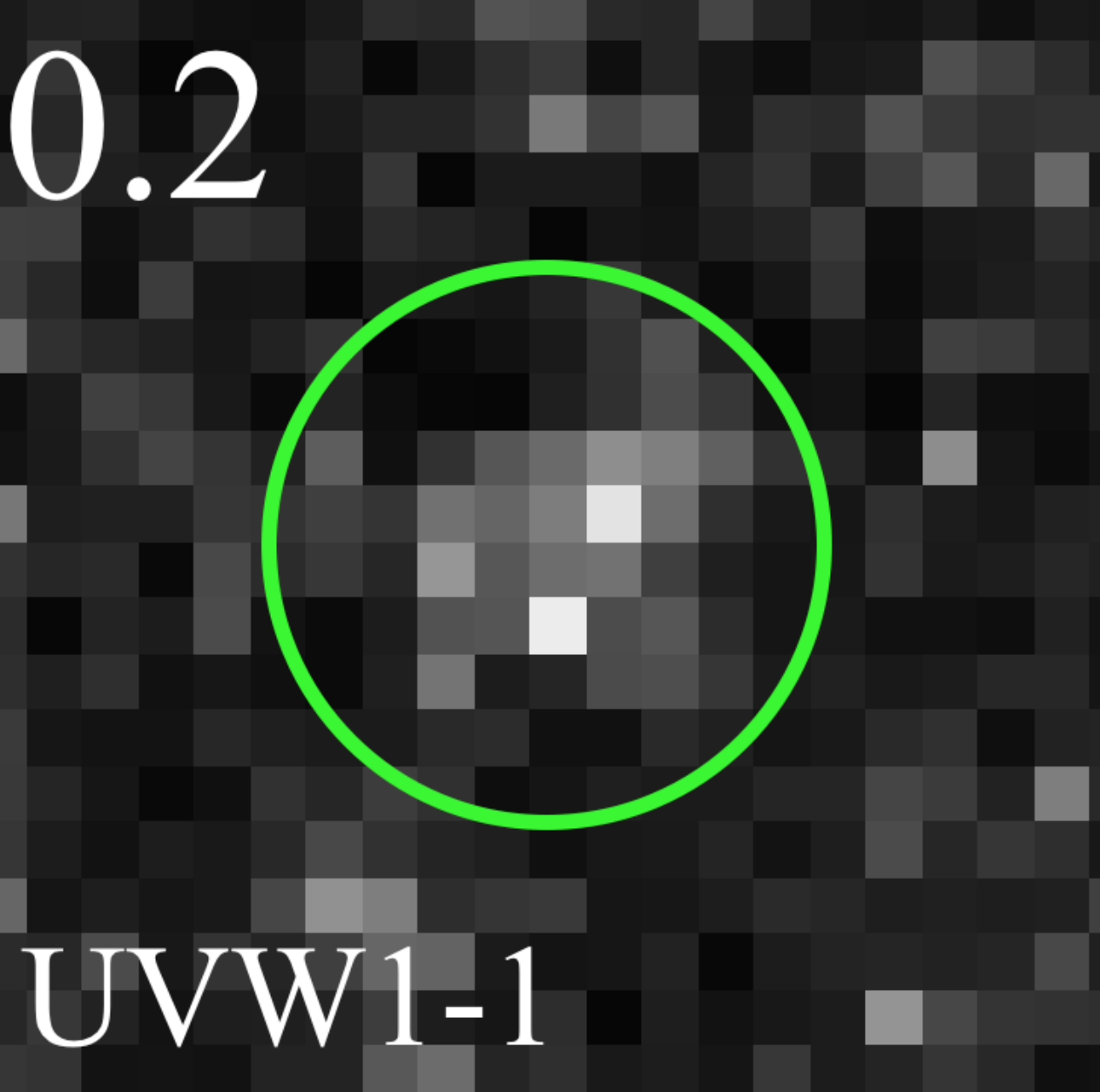} &
   \includegraphics[width=.23\textwidth]{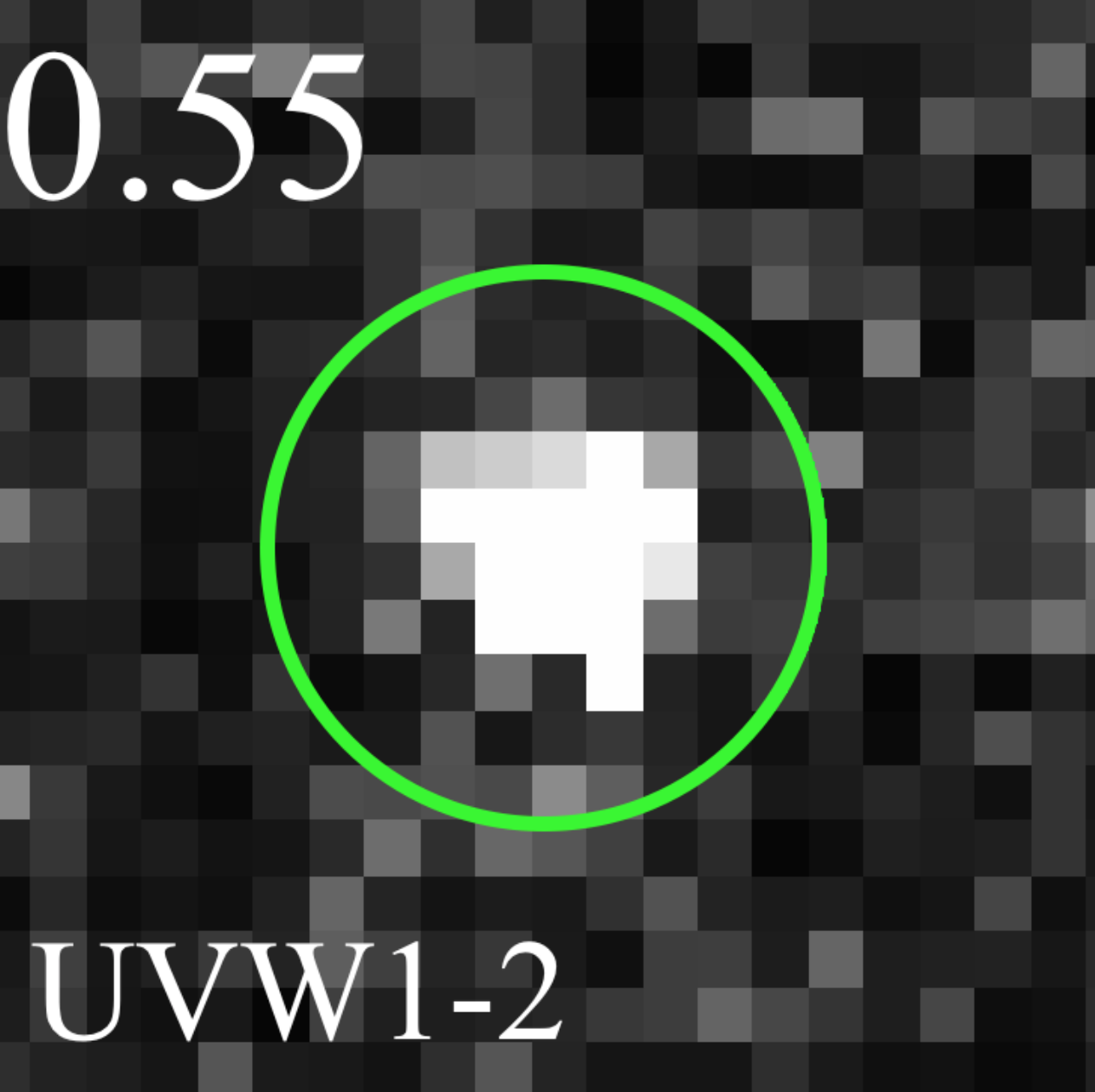} \\
  \includegraphics[width=.23\textwidth]{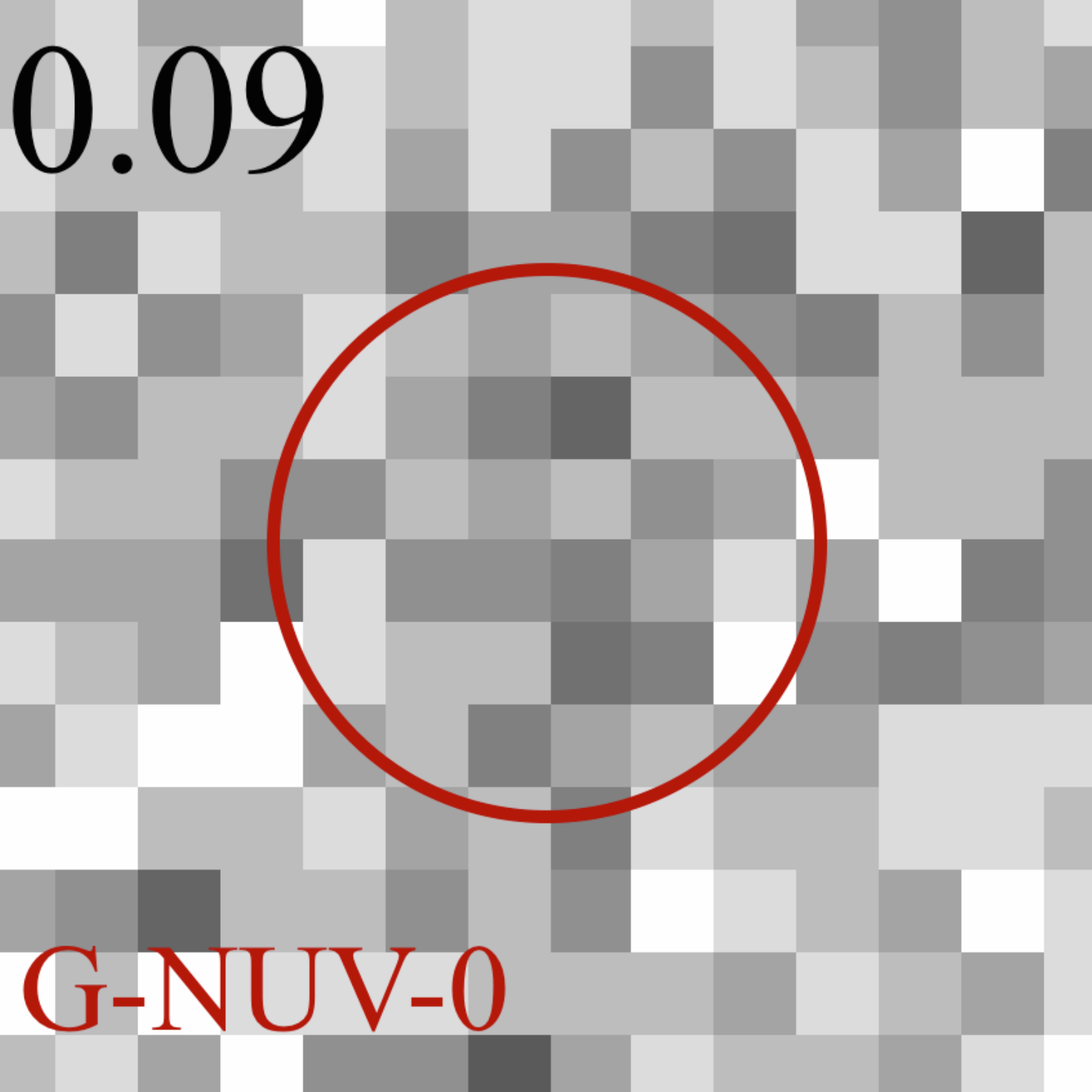} &
   \includegraphics[width=.23\textwidth]{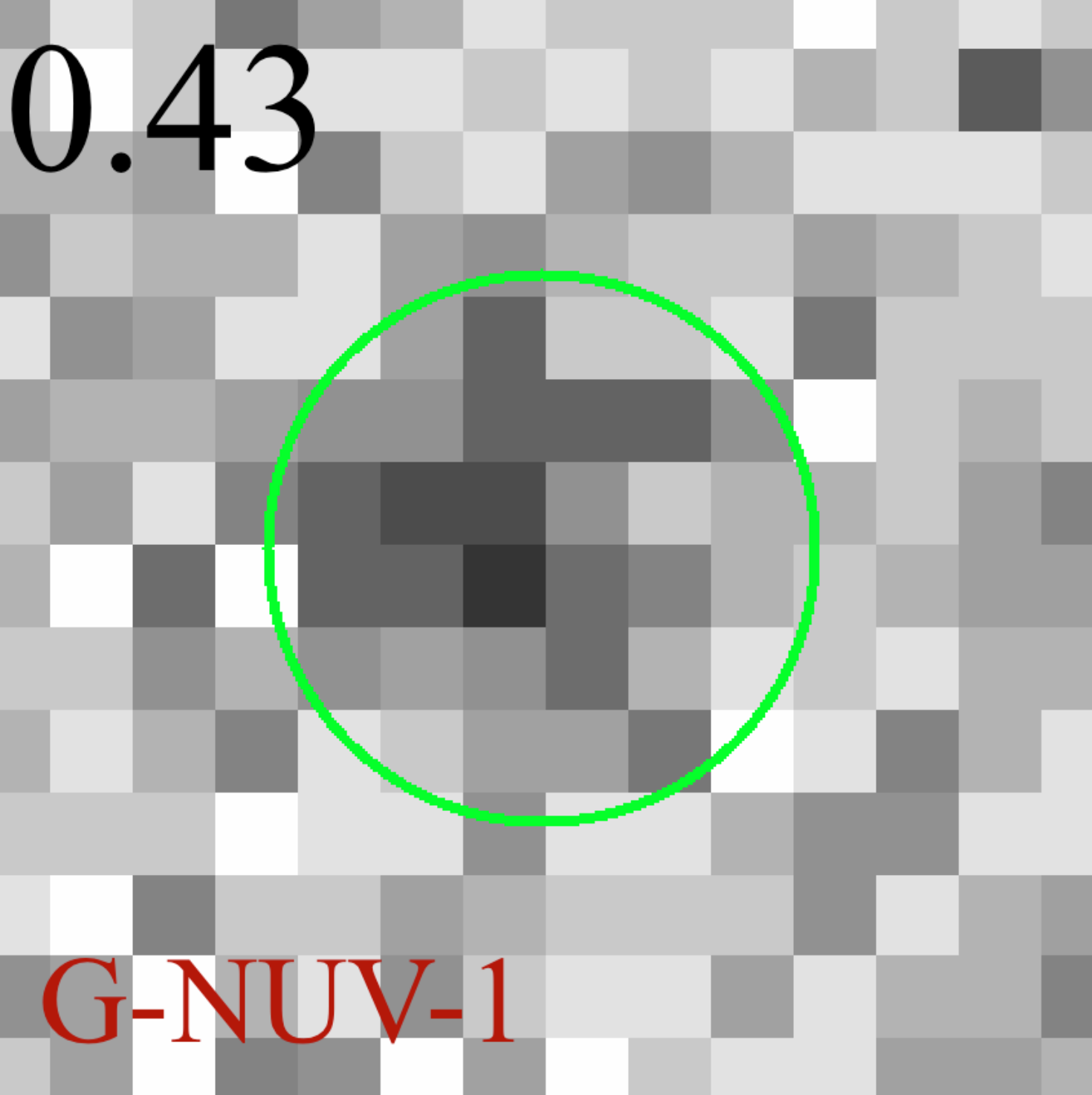} &
   \includegraphics[width=.23\textwidth]{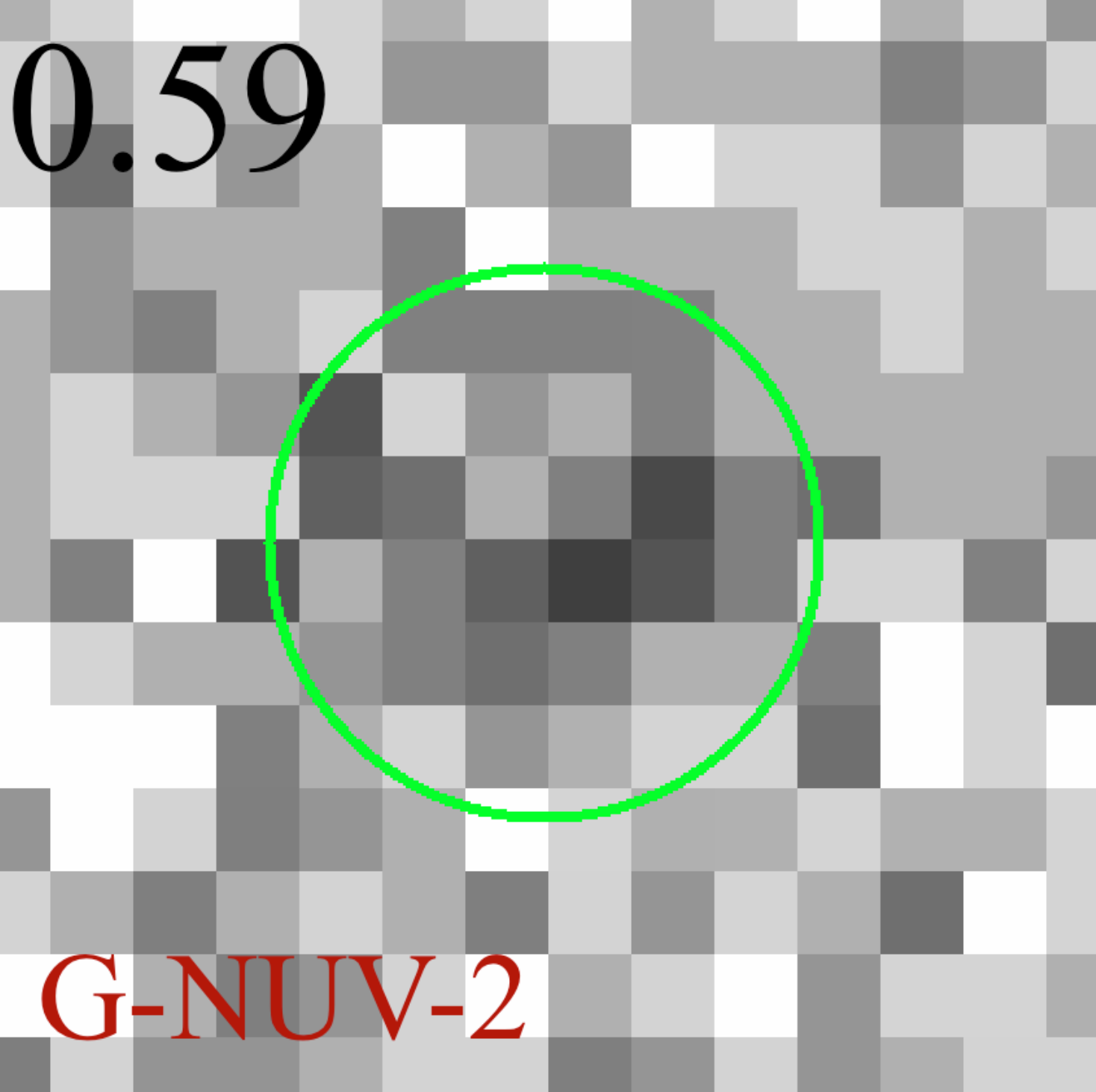} &
   \includegraphics[width=.23\textwidth]{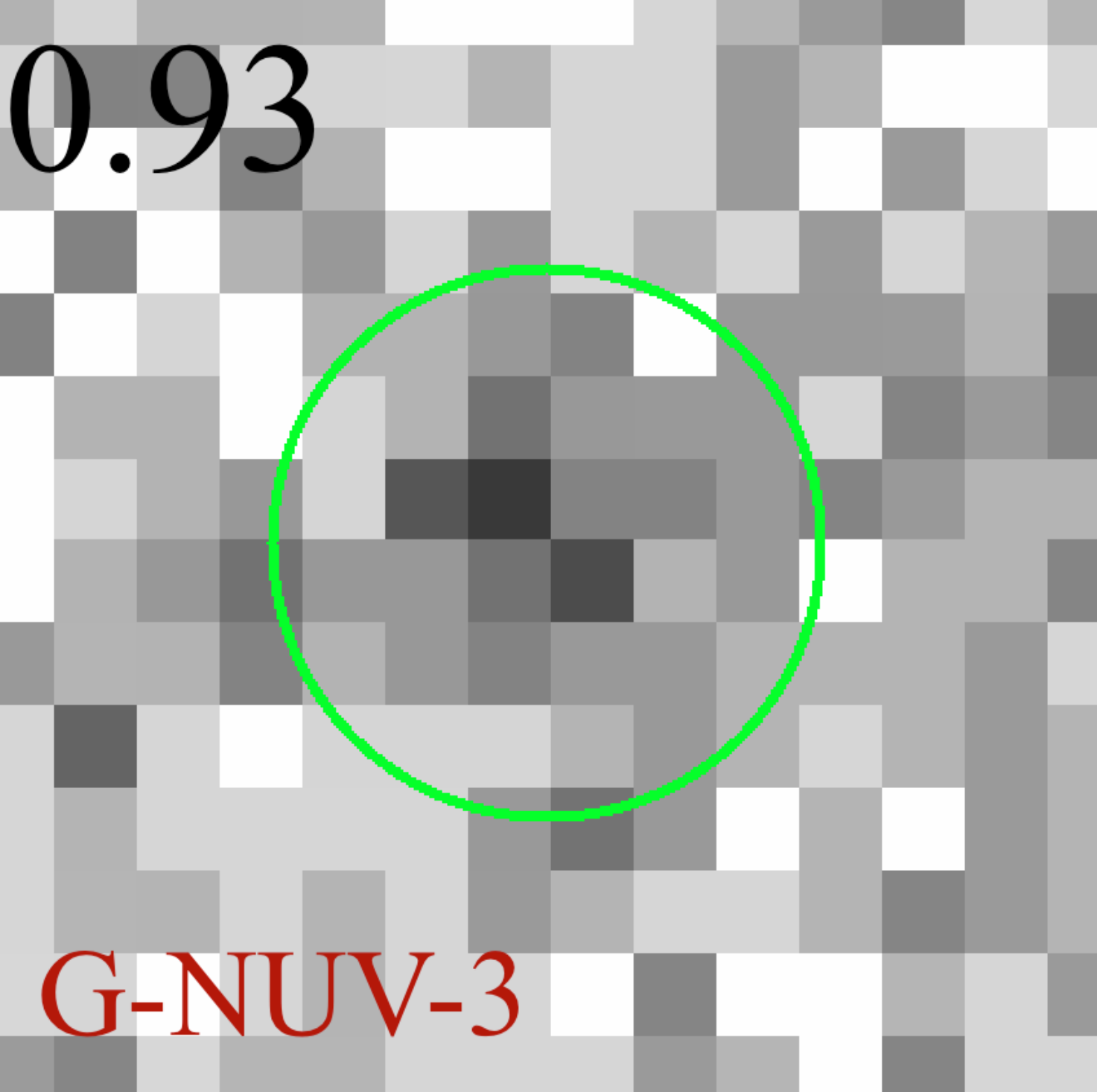} \\
  \end{tabular}
  \caption{MUV charts of PSR J1023+0038 during the pulsar state at different orbital phases. The filter and number of the image (see Table \ref{foto}) is indicated in the low-left corner. We have used G-NUV as abbreviation of \textit{GALEX}/NUV. The image G-NUV-0 (red circle) corresponds to a non-detection ($m_{MUV}>23$ mag) of the pulsar PSR J1023+0038 in 2009-03-07. The exposure time of 
that image is longer than the ones of G-NUV-2 and G-NUV-3. The orbital phase of the pulsar is indicated on the top-left corner.  Phase 0.25 corresponds to inferior conjunction of the companion. Each image is $20\arcsec \times 20 \arcsec$. Colors are inverted for the \textit{GALEX}/NUV images to get a better contrast.}
\label{finding}
\end{figure*}

%This mass is in agreement with masses of classical ``black widow'' systems which have masses $M_{bw}<0.1$ M$_{\sun}$ \citep{2013-Roberts}. 

\subsection{UV observations of XSS J12270$-$4859}

In Section \ref{parameters} we showed the observed large amplitude variations ($\sim0.9$ mags for F218W and $\sim0.6$ mags for F225W and F275W) in the folded light curves of our MUV observations in the covered phase ranges.
The modeled MUV light curves using \caps{icarus} and stellar atmospheres (Section \ref{parameters}) in a direct heating model, predict amplitude variations (between phases 0.25 and 0.75) of 
 $\sim$2.9, $\sim$2.4 and $\sim$1.8 mags for F218W, F225W and F275W. These modeled variations result independently of the assumption of the priors for all the parameters. In our three filters, the modeled MUV amplitude variations are larger than the variations observed by \citet{2015-martino} in the OM-U band ($\lambda_{eff}=3440$ \AA).
These authors observed variations of $\sim1.2$ mags and $\sim1.8$ mags in their observations 1 and 2 carried out in 2013 and 2014, respectively.
This suggests that, considering a direct heating model, we are more sensitive to trace the effect of heating on the companion star at our MUV wavelengths. 
Additionally, all our models predict that the companion has a temperature difference between the day-side and the night-side of $\sim1000$ K. 
The large observed amplitude variations in optical and estimated difference in temperatures, suggest that the donor star is highly irradiated by the NS, 
as also found by \citet{2015-martino}. 

Previously, we have also mentioned (Section \ref{parameters}) that the orbital parameters determined from optical data under the assumption of direct heating
fail to reproduce our MUV observations. This failure is independent of the priors assumed. 
There is always a deficiency in the filters F225W and F275W, 
whereas it could be an excess in the filter F218W.    
These results suggest that the MUV emission can not be explained by originating only from the heated companion star, instead an additional source of MUV radiation is required.  

We discard as the primary source of MUV light a hot NS (either heated because of the preceding accretion phase of the system, or due to hot spots on the NS magnetic pole) since in such scenario we would expect to see constant luminosity at all phases in the light curve, except at 0.25 when the companion would (partially) eclipse the NS (in the case of a high binary inclination). 
A strong MUV contribution from a residual accretion disc is as well unlikely since X-rays observations of XSS J12270$-$4859 
also show orbital variations \citep{2014bog, 2015-martino} that are consistent in phase with those in NUV \citep{2015-martino}. 
These X-ray observations show that the X-ray emission is predominantly non-thermal \citep{2014bog, 2014bassa, 2015-martino}. 
Given the long duration of the X-ray minima, it seems that the region which originates the X-ray emission is extended and possibly close to the companion star. 
This is consistent with X-ray emission coming from an intrabinary shock \citep[e.g.][]{2014bog,2014bassa,2015-martino, 2016-baglio} formed due to the interaction between the wind of the pulsar and 
matter outflowing from the companion star. Under that scenario seems unlikely that the matter from the companion star travels through the binary shock to form 
an accretion disc (or impact point) responsible for the MUV emission. 
Furthermore, optical observations of quiescent LMXBs show that these objects usually have $H{\alpha}$ excess \citep[][van den Berg et al., in prep.]{2004Haggard}.
This contrasts with the lack of $H{\alpha}$ emission in the spectrum of XSS J12270$-$4859 \citep{2014bassa}. 
Moreover, the SED (Figure \ref{bbsed}) does not show indications of a BB component in the MUV. 
 
Since the direct heating model also produces a poor fit on the optical observations 
(suggesting the presence of another component, Section \ref{parameters}),
it means that the observed optical fluxes could be a combination of emission from the companion star (larger contributor) and from the intrabinary shock.
This would make the companion to look brighter than in reality is. That scenario would help explaining the failure of the direct heating model on the optical observations, the overestimation by the model of the MUV flux in the bands F225W and F275W, and the underestimation of the flux in F218W. 
Another aspect is that the MUV amplitude variations seem to be correlated with the orbital phase, reaching the maximum at around 0.75 (superior conjunction of the companion), and the minimum at 0.25. Similar orbital variations are seen in the OM-U band observations of \citet{2015-martino}, which sample the whole orbital period and clearly show a maximum at phase 0.75 too. 
A possible scenario to understand why the peak is reached at phase 0.75 is that we see a larger emitting area from the presumable intrabinary shock, thus we see more UV radiation. At phase 0.25 the companion would partially obscure the shock, therefore the light curve would reach its minimum value. 

Assuming that the true folded light curve in the filter F225W is symmetric (see Section \ref{sed}),
the SED of XSS J12270$-$4859 suggests a possible excess in the filter F218W with respect to the BB model. 
At the same orbital phase, the data point in F225W approximately fits to the same model.  
Taking into account that the wavelength range of the filter F218W is completely embedded into the wavelength range of the filter F225W\footnote[13]{Transmission curves can be found at http://www.stsci.edu/hst/wfc3/ins\_performance/throughputs}, 
the excess in F218W could be due to the presence of emission lines in that band (probably coming from the intrabinary shock). The same lines would be also present in the filter F225W. However, it is possible that they might be smoothed out by the continuum emission in the broader F225W filter. Unfortunately, because our observations do not cover a whole orbital period in any of the three UV filters, 
we can not conclusively confirm whether the MUV light curves are indeed symmetric or not. 
Photometric MUV observations that cover the entire orbital period of the system are needed to investigate this further.  
Spectroscopic observations in the far-UV and MUV with a sensitive telescope like $HST$, would also help to confirm or discard the presence of emission lines. 
We also note that the data point in the filter F275W seems to show a deficiency with respect to the BB curve.
However, that data point was excluded from the fit because it corresponds to an orbital phase of 0.42 (which is the closest one to 0.52), 
when we assume that the real light curve in F275W is symmetric (see Section \ref{sed}). 
Thus, it is possible that the deficiency is explained by large amplitude variations in that band in a very short orbital phase range (note that the current observations in that filter show an amplitude variability of $\sim0.6$ mags in a phase range of $\sim0.1$).

\subsection{The possible configuration of PSR J1023+0038 in the MSP state}

We found that this tMSP shows MUV luminosity variations at different orbital phases during the pulsar state (See Figures \ref{GALEX_lc} and \ref{finding}). 
In the observations of 2009, the system was not detected at phase 0.09 (m$_{MUV} > 23$ mag) but it was detected at phase 0.43.
These phases roughly correspond to the passages of the ascending and descending nodes. 
Hence, we would expect the system to be detected in both phases (assuming symmetry). 
Since the time difference between these observations is only a couple of hours, the flux variations can not easily be explained by changes in the physical processes responsible of the UV radiation (as can be hypothesized to explain the observed variations between 2009 and 2010 at similar phases). 
A possible explanation for this behavior is the presence of an asymmetric intrabinary shock that emits MUV radiation anisotropically,
affecting the way in which that radiation is observed along the orbit. Such kind of binary shocks in black widow and redback systems have been studied by \citet{2016romani}. These authors show that the geometry of the intrabinary shock is controlled by the wind momentum flux 
ratio\footnote[14]{Defined as ${\dot M}vc/{\dot E}$, with ${\dot M}$ the mass-loss rate of the companion, $v$ the companion's wind speed, 
c the speed of light and ${\dot E}$ the pulsar spin-down power.} 
and the ratio between the companion's wind speed and the orbital velocity. Thus, variations in any of these parameters would lead to the formation of an asymmetric shock. These kind of models reproduce well the observed X-ray light curves of black widow and redback millisecond radio pulsars \citep{2016romani}. 
This means that such configuration could help explaining the observed asymmetric X-ray light curve of PSR J1023+0038 in its pulsar state \citep{2010archi,2011bog}. 
This curve is different from that of XSS J12270$-$4859 which seems to be symmetric. 
In fact, when we compare the four \textit{GALEX} data points to the folded X-ray light curve of PSR J1023+0038 \citep[Figure 2 in][]{2011bog}, we find that these observations are correlated. 
At phases 0.09-0.41, \citet{2011bog} find a minimum in the X-ray light curve. 
They attribute it to gradual occultation of an extended region (intrabinary shock) by the companion star. 
This range of phases corresponds to our non-detection ($m_{MUV}>23$ mag) of the system in the MUV at phase 0.09, but the detection of it at 0.43. \citet{2011bog} also find that the X-ray light curve reaches the maximum between phases 0.5-0.6, which matches with our brightest detection with \textit{GALEX} at phase 0.59.
Finally, our detection at phase 0.93 also corresponds to the detection of the system in X-rays at the same orbital phase \citep[approximately at the beginning of the radio eclipses;][]{2011bog}. These results strongly support the presence of an asymmetric intrabinary shock in PSR J1023+0038.  
Unfortunately, in the UV there are not enough observations of the system (in the pulsar state) to investigate this possibility further. We point out that the MUV predicted flux of the pulsar PSR J1023+0038 before late June 2013 by \citet{2014li} and \citet{2014takata} is of the order of $\sim10^{-12}$ erg cm$^{-2}$ s$^{-1}$ (from their Figures 5 and 3, respectively), which is a bit larger than the one we obtained 
using our \textit{GALEX} magnitudes for the same system. The MUV fluxes obtained go from $\sim5\times10^{-13}$ erg cm$^{-2}$ s$^{-1}$ to $< 7 \times 10^{-15}$ erg cm$^{-2}$ s$^{-1}$, depending on the orbital phase of the system.
New observations taken during a future MSP state of PSR J1023+0038 (if the object makes a transition again) will help to understand its UV luminosity behavior.

\subsection{Comparison of the MUV flux of XSS J12270$-$4859 and PSR J1023+0038}
 
In order to make a fair comparison of the MUV flux densities of both tMSPs, we have looked at data points with similar orbital phases using UV filters with similar effective wavelength.  
Therefore we compared the observation in the filter F218W ($\lambda_{eff}=2216$ \AA) for XSS J12270$-$4859 at an orbital phase of 0.59, to the observation of PSR J1023+0038 at the same orbital phase in
the \textit{GALEX} NUV filter ($\lambda_{eff}=2274$ \AA). We obtained a dereddened magnitude of $19.716 \pm0.057$ for XSS J12270$-$4859 which is equivalent to a flux density of $6.05 \pm 0.31 \times 10^{-17}$ erg s$^{-1}$ cm$^{-2}$ \AA$^{-1}$.
On the other hand, at the same orbital phase, we got a dereddened magnitude of $19.27\pm0.09$, equivalent to a flux density of $8.9 \pm 0.7 \times 10^{-17}$ erg s$^{-1}$ cm$^{-2}$ \AA$^{-1}$ \ for PSR J1023+0038. 
Considering a distance of $\sim1.4$ kpc to each pulsar, as determined from radio DM and parallax measurements \citep{2015roy, 2012deller}, 
we get a MUV luminosity of $3.15 \times 10^{31}$ erg s$^{-1}$ for XSS J12270$-$4859 and $4.75 \times 10^{31}$ erg s$^{-1}$ for PSR J1023+0038.
This means that the MUV luminosities for both systems are comparable, suggesting that similar physical processes occur in both tMSPs during the pulsar state. 
However, as we have seen, our folded light curve modeling might indicate a larger distance towards XSS J12270$-$4859 for M$_C> 0.01$ M$_{\sun}$. 
Therefore if we consider a distance to XSS J12270$-$4859 of 2.4 kpc or 3.2 kpc, as derived from our light curve fittings using MCMC,
the luminosities obtained are $9.5\times 10^{31}$ erg s$^{-1}$ and  $1.7 \times 10^{32}$ erg s$^{-1}$, respectively.
Thus, the tMSP PSR J1023+0038 is less bright in MUV. The difference could be due to changes in the geometry of the intrabinary shock between the two systems. 
More MUV observations during the pulsar phase for both objects would help to further investigate this scenario.

\section*{Acknowledgments}
The authors appreciate the comments of the anonymous referee, which improved this paper.
LERS acknowledges support from NOVA and a CONACyT (Mexico) fellowship. 
JVHS and ND are supported by a Vidi grant awarded to ND by the NWO. 
RW acknowledges support from a NWO top grant, Module 1. 
DA acknowledges support from the Royal Society. 
The authors thank Dr. D. de Martino who very kindly provided the g', r' and i' magnitudes used in this paper. 
They also thank Dr. Andrew Dolphin for his help with the use of the software DOLPHOT, Dr. Rene Breton for his input on the use of the \caps{icarus} software and Dr. Jason Hessels and Dr. Alessandro Patruno for useful conversations. 
The authors acknowledge the data provided by the ATNF Pulsar Catalogue. The NASA ADS abstract service was used to access to scientific publications and for getting the references used in this paper. 

\bibliographystyle{mn2e}
\bibliography{ref_tmsp}

\clearpage

\appendix

\begin{table*}
\section{Fitted phase folded light curves of XSS J12270$-$4859 }
\label{apen0}
\flushleft{
In this appendix we present the phase folded light curve fitting using MCMC in a direct heating model for observations in the filters $g', r'$ and $i'$ (left) and in the filters F218W, F225W and F275W (right). For the first set of plots we have assumed uniform priors with $0<q<20$, and for the second one Gaussian priors for $q$ but uniform priors for the other four parameters ($T_{base}$, $T_{irr}$, i and $\mu$). See Section \ref{parameters} for a more detailed description of the method.} 
\end{table*}

\begin{figure*}
\centering
\begin{tabular}{cc}
\large{Uniform priors with $0<q<20$ }\\
\vspace{0.1cm}
\twofigbox{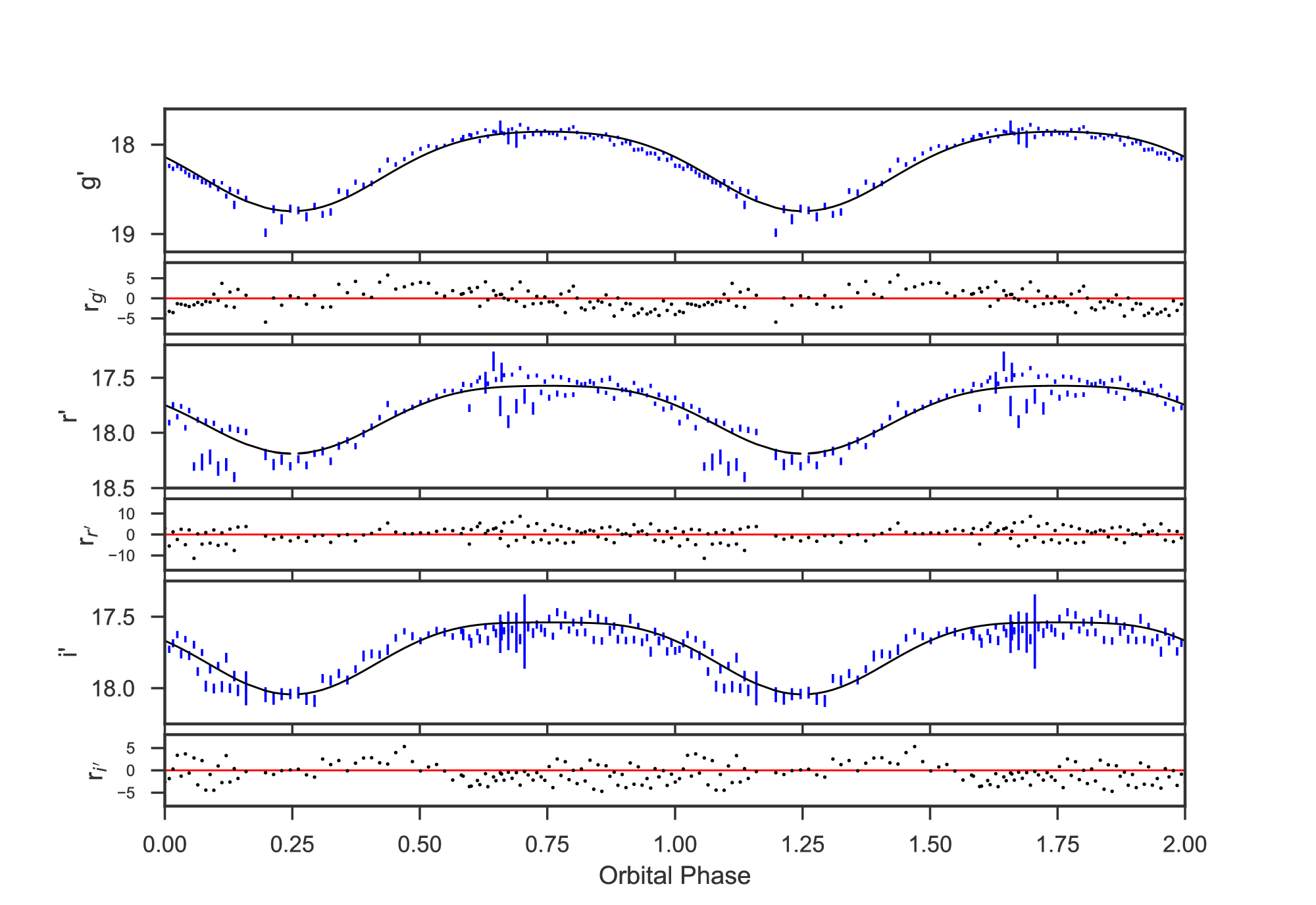}{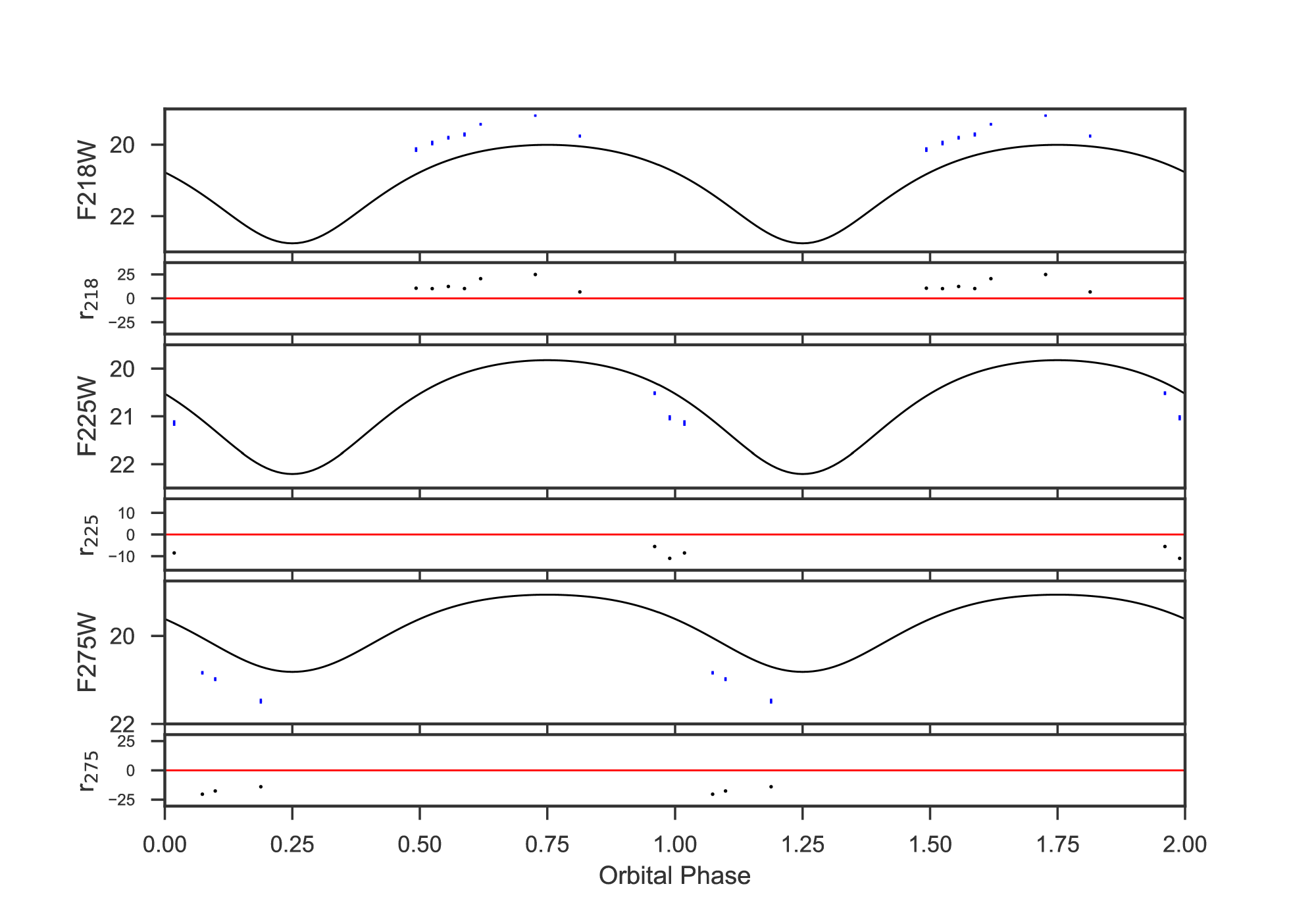}\\
\large{Gaussian prior for the mass ratio}\\
\twofigbox{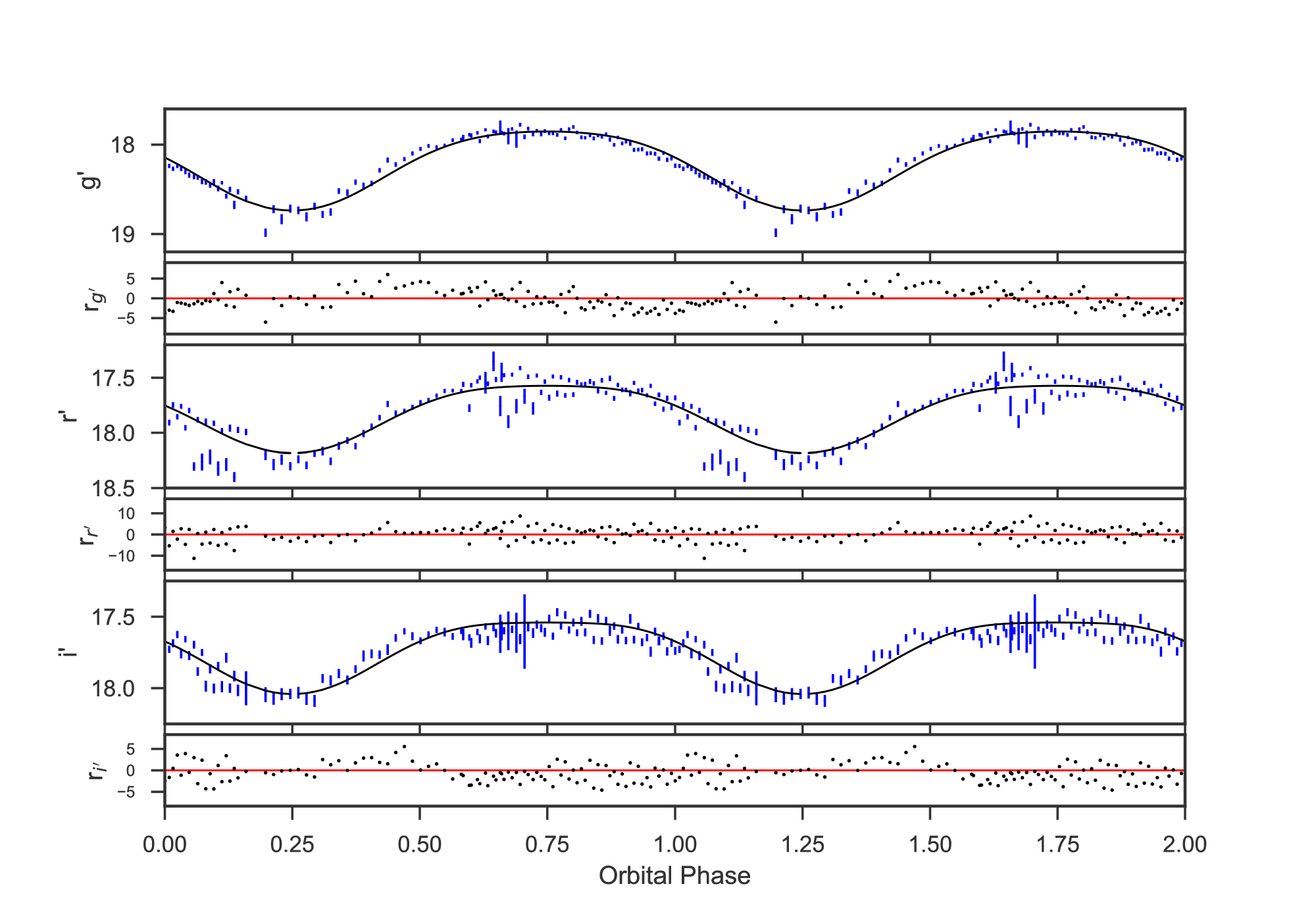}{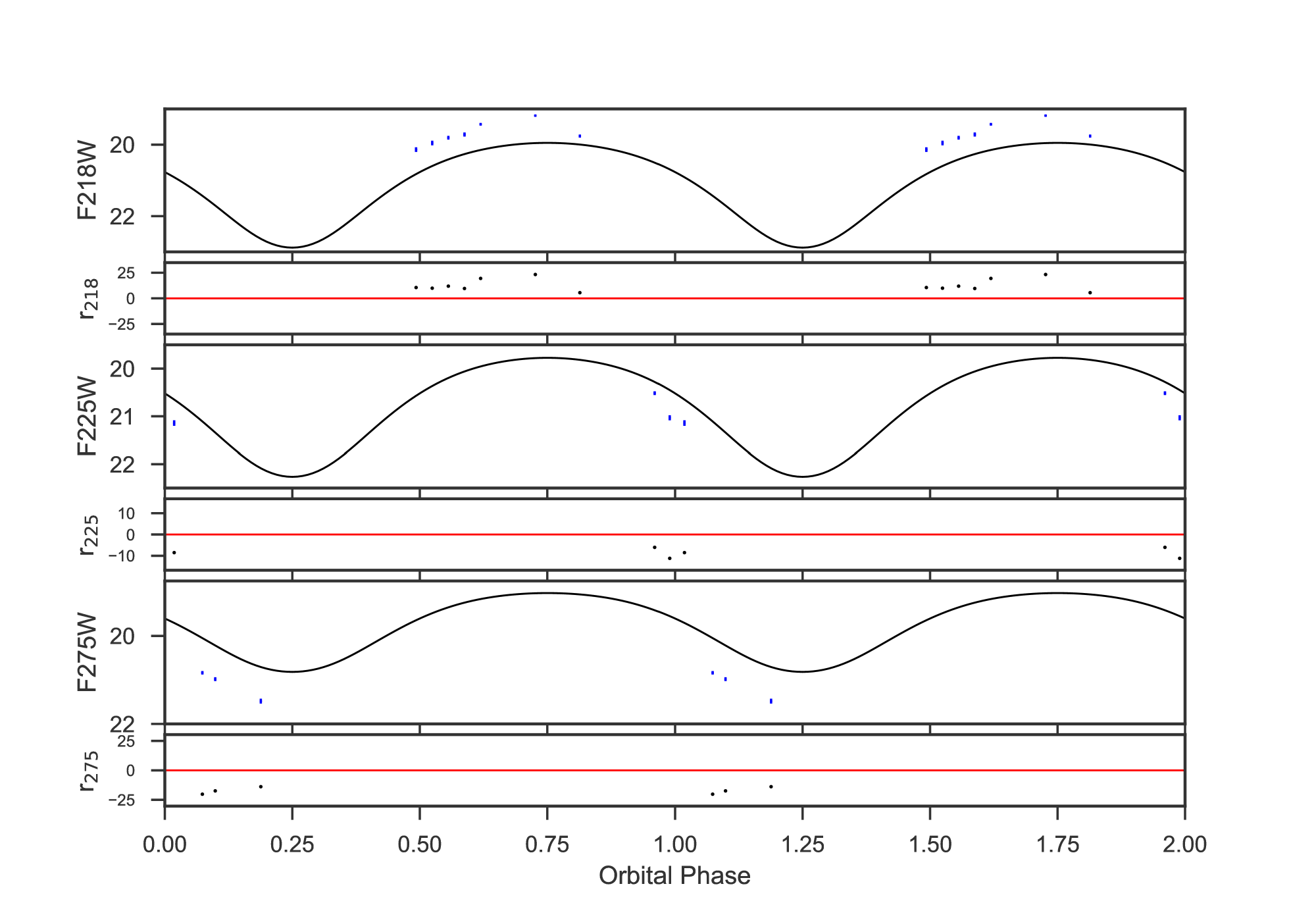}\\
\end{tabular}
 \caption{Phase folded light curves of XSS J12270$-$4859 in different filters. In the optical we have used only the measurements from January 2015 of \citet{2015-martino}. The data of February 2015 present very large scatter (see Section \ref{parameters}). The best fit is indicated with a solid gray line. 
Residuals (r) in each case are plotted below the corresponding light curves. 
These plots suggest the presence of systematic deviations in the system, which could affect the light curve fitting in these bands. 
Error bars represent the $1\sigma$ errors.
Phase 0.25 indicates that the companion is in inferior conjunction. 
Note how the modeled light curves in the filters $g', r'$ and $i'$ poorly fit to the data (${\chi}^2_{\nu}=7.3$ in both cases). 
Using the same system parameters, the models completely fail to reproduce the MUV observations. 
In both cases a deficiency with respect to the models in the filters F225W and F275W is observed, whereas in F218W there is an excess. 
For clarity, the phase is plotted twice. The data were barycentred and dereddened before creating the folded light curves.}
\label{lightcurvesappen}
\end{figure*}

\clearpage

\begin{table*}
\section{Posterior probability distributions}
\label{apen1}
\flushleft{
In this appendix we show the posterior probability distributions corresponding to the cases of uniform and Gaussian priors discussed in Section \ref{parameters}. Note that changes in the mass distribution imply changes in the distribution of luminosity, which in turn imply changes in the distribution of flux and distance to the system. In Table \ref{mcmc_par} we show that the larger the distance, the smaller the mass ratio. The obtained apparent magnitudes of the system help us
to determine the flux, which is related to its temperature. 
}
\end{table*}

\begin{figure*}
\centering
  \includegraphics[width=18cm]{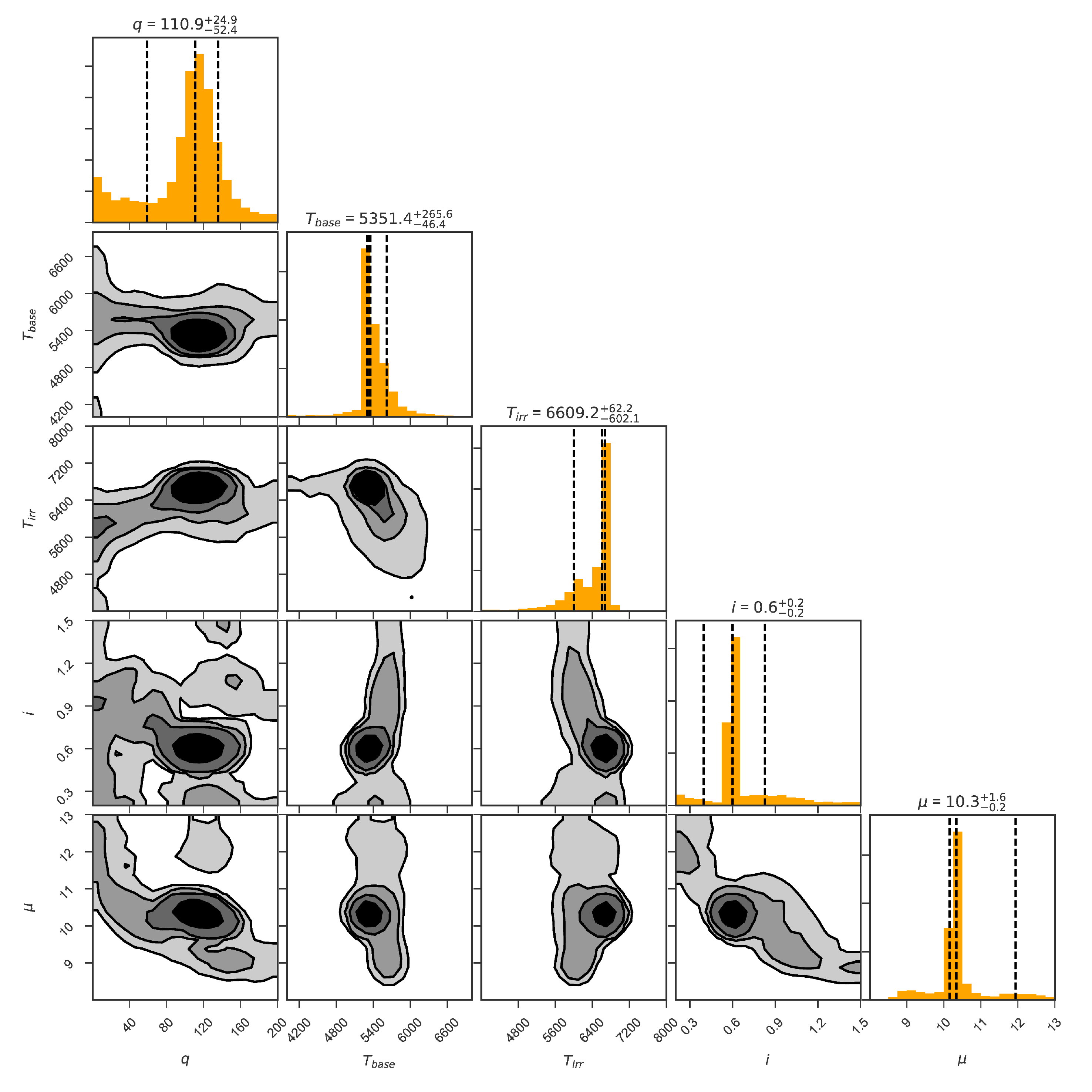} 
  \caption{Posterior probability distributions for the five parameters used to do the phase folded light curve fitting of XSS J12270$-$4859. 
We have assumed uniform priors for all the parameters and we used $0 < q < 200$. 
Histograms show the distribution of the probabilities for each parameter. The median value and the errors, corresponding to the 16\% and 84\% percentiles (vertical dashed lines) of the 1D distributions, are specified at the top of each column.
Contours show the probability for every combination of the respective parameters. 
The contours correspond to the 50\%, 68\%, 84\% and 95\% of the volume. Note the local and global maximum of the parameter $q$ at values $\sim11$ and 110.9, respectively. 
Units for $T_{base}$ and $T_{irr}$ are K, the units of i are radians. 
For this plot we have used the routine \textsc{corner}
\citep{citecorner}.}
 \label{uniform}
\end{figure*}

\begin{figure*}
\centering
  \includegraphics[width=18cm]{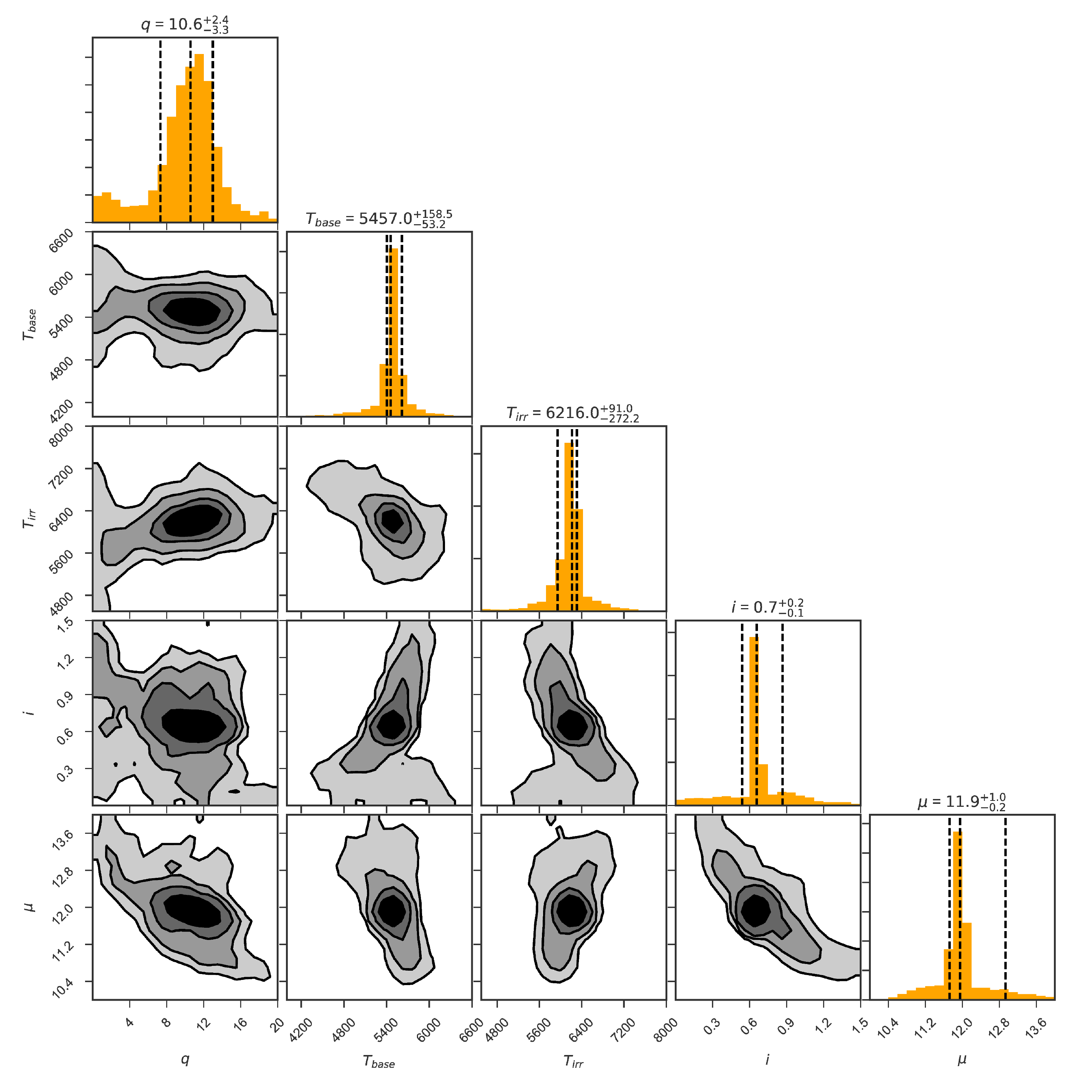} 
  \caption{Similar to Figure \ref{uniform} but with a domain of $q$ from 0 to 20.
Note the local and global maximum of the parameter $q$ at values $\sim3$ and 10.6, respectively (see Section \ref{parameters}).
Units for $T_{base}$ and $T_{irr}$ are K, the units of i are radians.  
For this plot we have used the routine \textsc{corner}
\citep{citecorner}.}
 \label{posteriors}
\end{figure*}

\begin{figure*}
\centering
  \includegraphics[width=18cm]{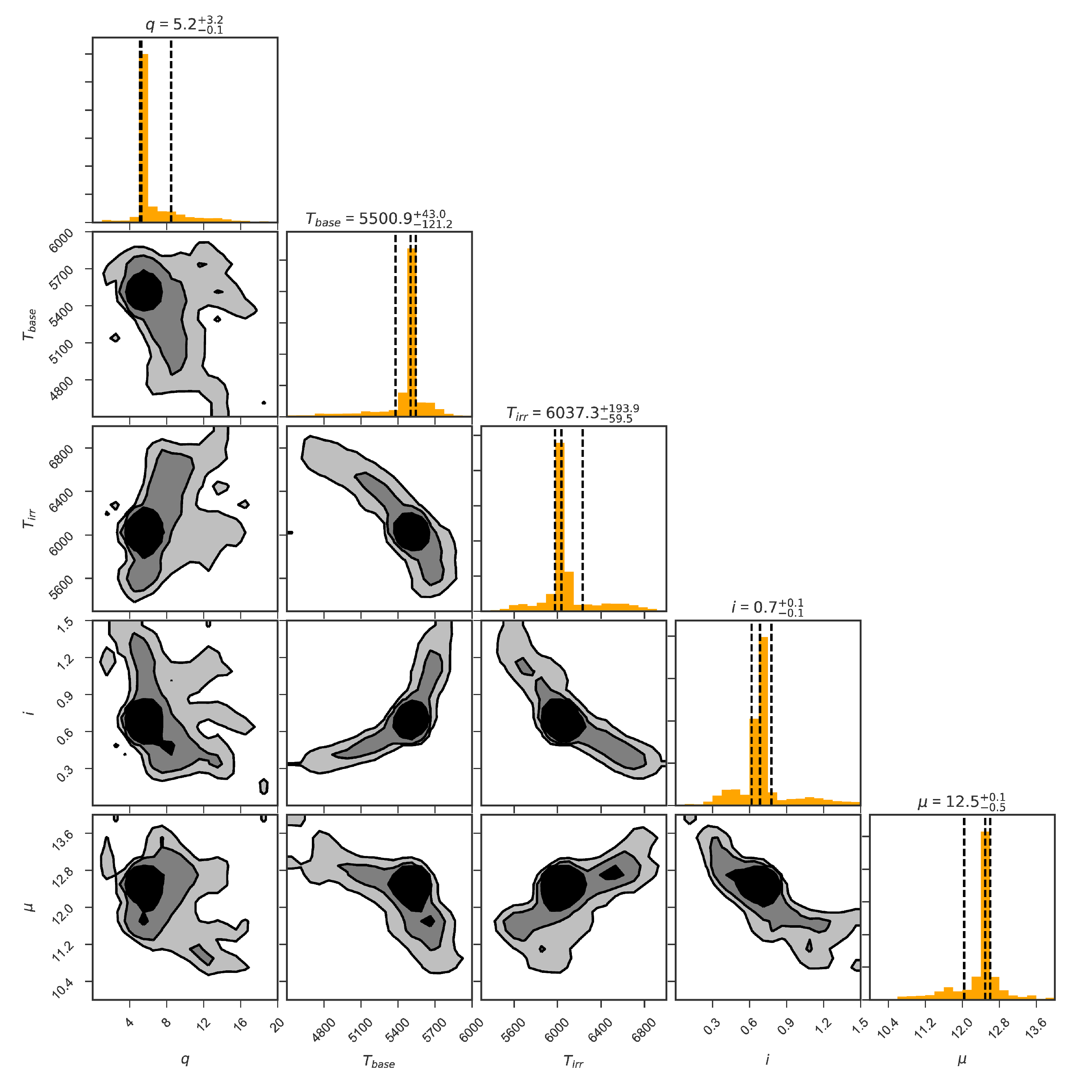} 
  \caption{Similar to Figure \ref{uniform} but using a Gaussian prior for the parameter $q$ and uniform priors for the other parameters. 
In this case contours correspond to the 68\%, 84\% and 95\% of the volume. 
Units for $T_{base}$ and $T_{irr}$ are K, the units of i are radians. 
For this plot we have used the routine \textsc{corner}
\citep{citecorner}.}
 \label{posteriors_gauss}
\end{figure*}

\end{document}